\documentclass[11pt]{article}
\usepackage{amssymb,amsmath}
\include{epsf}

\newcommand{\eqn}[1]{(\ref{#1})}
\def\appendix#1{\addtocounter{section}{1}\setcounter{equation}{0}
\renewcommand{\thesection}{\Alph{section}}
\section*{
\thesection\protect\indent \parbox[t]{11.715cm} {#1}}
\addcontentsline{toc}{section}{Appendix\thesection\ \ \ #1} }
\newcommand{\complex}{{\bb C}} 
\newcommand{\real}{{\bb R}} 
\newcommand{\id}{{1\!\!1}} 

\font\mybb=msbm10 at 12pt
\def\bb#1{\hbox{\mybb#1}}

\newcommand{\Tr}[1]{\:{\rm Tr}\,#1}

\def\be{\begin{equation}}
\def\ee{\end{equation}}
\def\bea{\begin{eqnarray}}
\def\eea{\end{eqnarray}}
\def\beq{\begin{equation}}
\def\eeq{\end{equation}}
\def\beqa{\begin{eqnarray}}
\def\eeqa{\end{eqnarray}}

\newcommand{\Ao}{\hat{A}}

\newcommand{\Do}{\hat{D}}

\newcommand{\No}{\hat{N}}

\newcommand{\Uo}{\hat{U}}

\newcommand{\Po}{\hat{P}}

\newcommand{\Go}{\hat{G}}

\newcommand{\Yo}{\hat{Y}}

\newcommand{\Fo}{\hat{F}}

\newcommand{\Lo}{\hat{L}}
\newcommand{\To}{\hat{T}}
\newcommand{\fo}{\hat{f}}
\newcommand{\go}{\hat{g}}
\newcommand{\baz}{\bar{z}}
\newcommand{\baw}{\bar{w}}

\newcommand{\ao}{\hat{a}}
\newcommand{\aod}{\hat{a}^{\dag}}

\newcommand{\bz}{\overline{z}}

\newcommand{\hil}{\mathcal{H}}

\newcommand{\la}{\langle}
\newcommand{\ra}{\rangle}
\newcommand{\del}{\partial}

\begin{document}
\begin{titlepage}
\begin{flushright}

\baselineskip=12pt DSF--4--2005\\ SISSA 37/2005/FM 
\end{flushright}

\begin{center}

\baselineskip=24pt

{\Large\bf The beat of a fuzzy drum:} \\ {\large\bf Fuzzy Bessel
functions for the disc}

\baselineskip=14pt

\vspace{1cm}

{{\bf Fedele~Lizzi, Patrizia~Vitale and Alessandro~Zampini$^{*}$}}
\\[6mm]
 {\it Dipartimento di Scienze Fisiche, Universit\`{a} di
Napoli {\sl Federico II}\\ and {\it INFN, Sezione di Napoli}\\
Monte S.~Angelo, via Cintia, 80126 Napoli, Italy}\\ {\tt
fedele.lizzi, patrizia.vitale, alessandro.zampini@na.infn.it}
\\ \vspace{1cm} 
$^{*}$ present address: {\it S.I.S.S.A. - Scuola Internazionale Superiore di Studi Avanzati -
\\
via Beirut 2-4, 34014 Trieste, Italy} \\ {\tt zampini@sissa.it}\\
[10mm]

\end{center}

\vskip 2 cm

\begin{abstract}
The fuzzy disc is a matrix approximation of the functions on a
disc which preserves rotational symmetry. In this paper we
introduce a basis for the algebra of functions on the fuzzy disc
in terms of the eigenfunctions of a properly defined fuzzy
Laplacian. In the commutative limit they tend to the
eigenfunctions of the ordinary Laplacian on the disc, i.e.\ Bessel
functions of the first kind, thus deserving the name of fuzzy
Bessel functions.
\end{abstract}

\end{titlepage}

\section{Introduction}
Noncommutative geometry~\cite{Connes,Landi,ticos,madorebook} has
suggested the introduction of \emph{fuzzy} spaces, namely the
approximation of the abelian algebra of functions on 
an ordinary space, with a finite rank
matrix algebra, which preserves the symmetries of the original
space, at the price of noncommutativity. The first formalization
of the idea of a fuzzy space was introduced by Madore
\cite{madorefs}, for the sphere. In his approach, a fuzzy sphere
is a sequence of nonabelian algebras, generated (loosely speaking)
by three ``noncommutative coordinates'' which satisfy $x_ix_i=1$
and $[x_i,x_j]=\kappa \varepsilon_{ijk} x_k$, with $\kappa$ some
noncommutativity parameter. The connection with the finite
dimensional representations of $SU(2)$ is immediate, and with
$\kappa$ proportional to the inverse of the rank of the
representation, one can think of recovering functions on the
sphere in some limit. This limit has been made rigorous by
Rieffel~\cite{rieffelGHdist} at the level of topological ``quantum
spaces'', but here we are more interested in the approximations
that the fuzzy matrix geometry can give to the space of functions
(and hence of fields) defined on some space. This is efficiently
done defining ``fuzzy spherical
harmonics''~\cite{ChuMadoreSteinacker,Ramgoolam}, a basis for the
matrix algebra, which in the limit converges to the ordinary
spherical harmonics. Field theory on the fuzzy sphere is a very
active field of research. A partial list of references
is~\cite{fuzzysphfield}.

After the fuzzy sphere, similar approximations for higher
dimensional spheres~\cite{higherspheres}, for the torus, based on the
noncommutative torus algebra~\cite{rieffelFT,portoroz}, and for
projective spaces~\cite{projective}, have also been studied. In
previous work~\cite{beatles,balfest}, we proposed a fuzzy
approximations for the algebra of functions on a disc (see
also~\cite{BalSeckinKumar,albalfest}). It is the first example of
a fuzzy approximation for a space that classically has a boundary. The
space is defined through a suitable truncation of the
noncommutative plane. In this paper we examine the spectral
properties of the fuzzy Laplacian, finding its eigenvalues and
eigenvectors, and comparing them with the exact case and give a
direct way to associate, to functions on the disc, their fuzzy
counterpart, i.e.\ to approximate the abelian algebra of functions
by a sequence of finite rank matrices. We introduce a basis for
the algebra of functions on the fuzzy disc in terms of symbols of
the eigenmatrices of the fuzzy approximation to the Laplacian.
These symbols are seen to converge, as the rank of the matrices
increases, towards a set of functions on the plane which coincide
with the ordinary Bessel functions of first kind (eigenfunctions
of the ordinary Laplacian with Dirichlet boundary conditions) for
points on a plane inside a disc, and to zero for points outside
this disc. These are given the name of fuzzy Bessel functions.

The paper is organised as follows. In the next section we review
the fuzzy sphere from a perspective which will render easy the
generalization to the disc case. The main tool we use is the
Weyl-Wigner formalism, and the concept of generalized coherent
sates, which we briefly review in appendices~\ref{appWW}
and~\ref{appz} respectively. In section~\ref{se:disc} we review the
fuzzy disc.  A Weyl-Wigner isomorphism is introduced in terms of
functions on the plane, where noncommutativity is represented by a
parameter $\theta$. In this formalization, there is no natural
concept of a sequence of finite dimensional Hilbert spaces, or
finite rank matrix algebras, therefore a  truncation in the
algebra of operators is introduced, with respect to a specific
basis in the Hilbert space. Constraining the dimension $N$ of the
truncation to the noncommutativity parameter, $N\theta=R^{2}$, one
obtains then a sequence of finite rank matrices, converging
towards an abelian algebra of operators, that approximates
functions whose support is concentrated on a disc of radius $R$.
On this sequence of states, fuzzy derivatives and a fuzzy
Laplacian are defined and the spectral properties of the latter
are studied. Then in section~\ref{se:fuzzybessel} we study
eigenvalues and eigenvectors and show that they (rather quickly)
converge to the ordinary Bessel functions.

\section{The fuzzy sphere in the Weyl-Wigner
formalism}\label{sec:fs} \setcounter{equation}{0}

In this section we present the fuzzy sphere in the general context
of the Weyl-Wigner formalism
\cite{quantforsphere1,quantforsphere2,zampthesis}. We first set up
an isomorphism between a space of operators and a space of
functions on a sphere. Since a sphere is the coadjoint orbit of
the group $SU\left(2\right)$, the basic tool to introduce this map
is a system of coherent states, specialising the general arguments
of appendix~\ref{appz}.

To define a system of coherent states consider the unitary
irreducible representations for the group $SU\left(2\right)$. On
each finite dimensional Hilbert space $\complex^{N}$, with
$N=2L+1$ -here $\left(L=0,1/2,1,3/2\ldots\right)$-, a basis is given
by vectors that, in ket notation, are represented as $\left|L,M\right.\ra$
-with $M=\left(-L,-L+1,\ldots,L-1,L\right)$-. One has:
\beq
u\,\in\,SU\left(2\right)\,\stackrel{\hat{R}^{\left(L\right)}}{\mapsto}\,
\mathcal{B}\left(\complex^{N}\right) \ .
\eeq
The matrix elements of
this representation are given by:
\beq
\la
L,M|\hat{R}^{\left(L\right)}\left(u\right)|
L,M^{\prime}\ra=D_{MM^{\prime}}^{L}\left(u\right) \ .
\eeq
These are called Wigner functions~\cite{QTAngMom}.

The second step is to fix a fiducial state. One can choose the
highest weight in the representation: $|\psi_{0}\ra=| L,L\ra$. If
the group manifold is parametrised by Euler angles, then $u$
represents a point whose ``coordinates'' range through
$\,\alpha\in\left[\left.0,4\pi\right)\right.\,$,$\,\beta\in\left[\left.0,\pi\right)\right.\,$,
$\,\gamma\in\left[\left.0,2\pi\right)\right.\,$. Fixed the
fiducial vector, its stability subgroup $H_{\psi_{0}}$ by the
$\hat{R}^{\left(L\right)}$ representation is made by elements for
which $\beta=0$ (this condition is seen to be valid independently
of $N$). Two elements $u$ and $u^{\prime}$ are equivalent if
$u^{\dagger}u^{\prime}\in H_{\psi_{0}}$. It is possible to prove
that:
\beq 
SU\left(2\right)/H_{\psi_{0}}\approx S^{2} 
\eeq
identifying $\theta=\beta$ and $\varphi=\alpha\mod\,2\pi$. Chosen
a representative element $\tilde{u}$  in each equivalence class of
the quotient, the set of coherent states is defined as:
\beq
|\theta,\varphi,N\ra=\hat{R}^{\left(L\right)}\left(\tilde{u}\right)|
L,L\ra \ .
\eeq
The left hand side ket now explicitly depends on $N$,
the dimension of the space on which the representation takes
place. Projecting on the basis elements, one has:
\beqa
\la L,M| \theta,\varphi,N\ra &=& D_{ML}^{L}\left(\tilde{u}\right)
\ ,
\\ |\theta,\varphi,N\ra &=& \sum_{M=-L}^{L}
\left({\textstyle
\frac{\left(2L\right)!}{\left(L+M\right)!\left(L-M\right)!}}\right)^{1/2}
\left(\cos\theta/2\right)^{L+M}\left(\sin\theta/2\right)^{L-M}e^{-i\varphi
M}\,| L,M\ra \ . \nonumber
\eeqa
This set of states is nonorthogonal, and overcomplete
($d\Omega=d\varphi\,\sin\theta\,d\theta$):
\beqa
\la\theta^{\prime},\varphi^{\prime},N|\theta,\varphi,N\ra &=&
e^{-iL\left(\varphi^{\prime}-\varphi\right)}\,
\left[e^{i\left(\varphi^{\prime}-\varphi\right)}\cos \theta/2\cos
\theta^{\prime}/2 +\sin \theta/2\sin\theta^{\prime}/2\right]^{2L}
\ , \nonumber
\\
\id&=&\frac{2L+1}{4\pi}\int_{S^{2}}\,d\Omega|\theta,\varphi,N\ra\la\theta,\varphi,N|
 \ . \label{SU2complete}
\eeqa
Using this set of vectors it is possible to define a map, from the
space of operators on a finite dimensional Hilbert space to the
space of functions on the sphere $S^{2}$:
\beqa
\Ao^{\left(N\right)}\,\in\,\mathcal{B}\left(\complex^{N}\right)\,\approx\,
\mathbb{M}_{N}\,\,\,&\mapsto&\,\,\,A^{\left(N\right)}
\,\in\,\mathcal{F}\left(S^{2}\right) \ , \nonumber
\\
A^{\left(N\right)}\left(\theta,\varphi\right)\,&=
&\,\la\theta,\varphi,N|\Ao^{\left(N\right)}|\theta,\varphi,N\ra \
. \label{berezinsymsph}
\eeqa
This is a way to map every finite rank matrix into a function on a
sphere, called the Berezin symbol of the given matrix~\cite{berezinmap}. Among these
operators, there are $\Yo^{\left(N\right)}_{JM}$ whose symbols are
the spherical harmonics, up to order $2L$ (here $J=0,1,\ldots,2L$
and $M=-J,\ldots,+J$):
\beq
\la\theta,\varphi,N|\Yo^{\left(N\right)}_{JM}|\theta,\varphi,N\ra\,=
\,Y_{JM}\left(\theta,\varphi\right) \ ,
\eeq
these operators are called \emph{fuzzy harmonics}. The origin of
this name must be traced back to the definition, on each finite
rank matrix algebra $\mathbb{M}_{N}$, of an operator, in terms of
the generators $\Lo_{a}^{\left(N\right)}$:
\beq
\left[\hat{L}_{a}^{\left(N\right)},\hat{L}_{b}^{\left(N\right)}\right]\,=
\,i\epsilon_{abc}\hat{L}_{c}^{\left(N\right)} \ , \label{su2alg}
\eeq
representing the Lie algebra of the group $SU\left(2\right)$ on
the space $\complex^{N}$:
\beqa
\nabla^{2}\,&:&\,\mathbb{M}_{N}\,\mapsto\,\mathbb{M}_{N} \ ,
\nonumber
\\
\nabla^{2}\Ao^{\left(N\right)}\,&=&\,\left[\Lo_{s}^{\left(N\right)},
\left[\Lo_{s}^{\left(N\right)}, \Ao^{\left(N\right)}\right]\right]
\ .
\eeqa
This operator is called the \emph{fuzzy Laplacian}. Its spectrum
is given by eigenvalues $\mathcal{L}_{j}=j\left(j+1\right)$, where
$j=0,\ldots,2L$, and every eigenvalue has a multiplicity of
$2j+1$. The spectrum of the fuzzy Laplacian thus coincides, up to
order $2L$, with the one of its continuum counterpart acting on the
space of functions on a sphere. The cut-off of this spectrum is of
course related to the dimension of the rank of the matrix algebra
under analysis. The fuzzy harmonics are the eigenstates of this
operator.

Fuzzy harmonics are a basis in each space of matrices
$\mathbb{M}_{N}$. They are orthogonal but non normalized and are
proportional to the polarization operators
$\To^{\left(N\right)}_{JM}$ defined in terms of the Clebsh-Gordan
coefficients for $SU(2)$~\cite{QTAngMom}:
\beqa
\left[\Lo_{\mu}^{\left(N\right)},\To^{\left(N\right)}_{JM}\right]\,&=&\,
\sqrt{L\left(L+1\right)}\,C^{L\,M+\mu}_{L\,M\
1\,\mu}\,\To^{\left(N\right)}_{LM+\mu} \ , \nonumber
\\
Tr\left[\To^{\left(N\right)\,\dagger}_{JM}\,
\To^{\left(N\right)}_{J^{\prime}M^{\prime}}\right]
&=&\delta_{JJ^{\prime}}\delta_{MM^{\prime}}\ , \nonumber
\\
\To^{\left(N\right)\,\dagger}_{JM}
&=&\left(-1\right)^{M}\To^{\left(N\right)}_{J-M} \ .
\eeqa
the proportionality being:
\beq
\hat{Y}^{\left(N\right)}_{JM} = \frac{J!}{\left(2lJ\right)^{J}}
\left[\frac{\left(2l+J+1\right)!}{4\pi\left(2l-J\right)!}\right]^{1/2}
\hat{T}^{\left(N\right)}_{JM} \ ,
\eeq
from which:
\beq
\Tr\left[\hat{Y}^{\left(N\right)\,\dagger}_{JM}
\hat{Y}^{\left(N\right)}_{J^{\prime}M^{\prime}}\right] =
\left(\frac{J!}{\left(2lJ\right)^{J}}\right)^{2}
\left(\frac{\left(2l+J+1\right)!}{4\pi\left(2l-J\right)!}\right)
\delta_{JJ^{\prime}}\delta_{MM^{\prime}} \ .
\eeq
Using this basis, an element $\Fo^{\left(N\right)}$ belonging to
$\mathbb{M}_{N}$ can be expanded as:
\beq
\Fo^{\left(N\right)}= \sum_{J=0}^{2L}\sum_{M=-J}^{J}\,
F^{\left(N\right)}_{JM}\,\Yo^{\left(N\right)}_{JM} \ ,
\eeq
with coefficients
\beq F^{\left(N\right)}_{JM}=\frac{
\Tr\left[\Yo^{\left(N\right)\dagger}_{JM}\,\Fo^{\left(N\right)}\right]}
{\Tr\Yo^{\left(N\right)\dagger}_{JM}\Yo^{\left(N\right)}_{JM}} \ .
\eeq
A Weyl-Wigner map can be defined simply mapping spherical
harmonics into fuzzy harmonics:
\beq
\Yo^{\left(N\right)}_{JM}\,\,\Leftrightarrow\,\,Y_{JM}
\left(\theta,\varphi\right) \ .
\eeq
This map clearly depends on the dimension $N$ of the space on
which fuzzy harmonics are realized. It can be linearly extended
 by: \beq \hat{F}^{\left(N\right)}=\sum_{J=0}^{2L}\sum_{M=-J}^{J}\,
F^{\left(N\right)}_{JM}\hat{Y}^{\left(N\right)}_{JM}\,\,\,\,\leftrightarrow\,\,\,\,
F^{\left(N\right)}\left(\theta,\phi\right)=\sum_{J=0}^{2L}\sum_{M=-J}^{+J}F_{JM}^
{\left(N\right)}Y_{JM}\left(\theta,\phi\right).
\label{WWmapsphere}\eeq This is a Weyl-Wigner isomorphism and it
can be used to define a fuzzy sphere. Given a function on a
sphere, if it is square integrable with respect to the standard
measure $d\Omega$, then it can be
expanded in the basis of spherical harmonics: \beq
f\left(\theta,\varphi\right)= \sum_{J=0}^{\infty}\sum_{M=-J}^{J}
\,f_{JM}\,Y_{JM}\left(\theta,\varphi\right).\label{harmexp} \eeq
Now consider the set of ``truncated" functions: \beq
f^{\left(N\right)}\left(\theta,\varphi\right)=
\sum_{J=0}^{2L}\sum_{M=-J}^{J}
\,f_{JM}\,Y_{JM}\left(\theta,\varphi\right) \ .
\label{fuzzharmexp}
\eeq
This is a vector space, but it is no more an algebra, with the
standard definition of sum and pointwise product of two functions,
as the product of two spherical harmonics of order say $2L$ has
spherical components of order larger  than $2L$. We have indeed:
\beq
\left(Y_{J^{\prime}M^{\prime}}Y_{J^{\prime\prime}
M^{\prime\prime}}\right)\left(\theta,\varphi\right) =
\sum_{J=\left|J^{\prime}-J^{\prime\prime}\right|}^{J
=\left|J^{\prime}+J^{\prime\prime}\right|}
\sum_{M=-J}^{J}\sqrt{\frac{\left(2J^{\prime}+1\right)
\left(2J^{\prime\prime}+1\right)}{4\pi\left(2J+1\right)}}
\,C^{J0}_{J^{\prime}0\,J^{\prime\prime}0}\,
C^{JM}_{J^{\prime}M^{\prime}\,J^{\prime\prime}M^{\prime\prime}}\,
Y_{JM}\left(\theta,\varphi\right)
\eeq 
in terms of Clebsh-Gordan coefficients for
$SU\left(2\right)$. If these truncated functions are mapped, via
the Weyl-Wigner procedure, into matrices:
\beq
f^{\left(N\right)}\left(\theta,\varphi\right)\,\mapsto\,\hat{f}^{\left(N\right)}
=\sum_{J=0}^{2L}\sum_{M=-J}^{+J}\,f_{JM}\hat{Y}^{\left(N\right)}_{JM}
\ , \label{fsmap}
\eeq 
then this set of truncated functions is given
the vector space structure of $\mathbb{M}_{N}$. But
$\mathbb{M}_{N}$ is even a \emph{nonabelian} algebra.
Invertibility of this association~(\ref{fsmap}) enables to define,
in the set of the truncated functions on the sphere, a non abelian
product, isomorphic to that of matrices:
\beq
\left(f^{\left(N\right)}*g^{\left(N\right)}\right)\left(\theta,\varphi\right)=
\sum_{J=0}^{2L}\,\sum_{M=-J}^{+J}\,\frac{Tr\left[\fo^{\left(N\right)}
\go^{\left(N\right)}
\Yo_{JM}^{\left(N\right)\,\dagger}\right]\,Y_{JM}\left(
\theta,\varphi\right)}
{\Tr\Yo^{\left(N\right)\dagger}_{JM}\Yo^{\left(N\right)}_{JM}} \ .
\eeq
The Weyl-Wigner map~(\ref{WWmapsphere}) has been used to make each
set of truncated functions a non abelian algebra
$\mathcal{A}^{\left(N\right)}\left(S^{2},*\right)$, isomorphic to
$\mathbb{M}_{N}$. The product of two fuzzy harmonics can be
obtained in term of the 6j-symbols from the product of two
polarization operators~\cite{ChuMadoreSteinacker}
\be
\To^{\left(N\right)}_{J^{\prime}M^{\prime}}
\To^{\left(N\right)}_{J^{\prime\prime}M^{\prime\prime}}=
\sum_{J}\left(-1\right)^{2L+J}\sqrt{\left(2J^{\prime}+1\right)
\left(2J^{\prime\prime}+1\right)} \left\{
\begin{array}
{ccc}{\scriptstyle J^{\prime}} & {\scriptstyle J^{\prime\prime}} & {\scriptstyle J} \\
{\scriptstyle L} & {\scriptstyle L} & {\scriptstyle L}
\end{array} \right\}
C^{JM}_{J^{\prime}M^{\prime}\,J^{\prime\prime}M^{\prime\prime}}
\hat{T}^{\left(N\right)}_{JM} \ .
\ee

These algebras can be seen as formally generated by matrices which
are the images of the norm $1$ vectors in $\real^{3}$, i.e.\
points on a sphere. They are mapped into multiples of the
generators $\Lo_{a}^{\left(N\right)}$ of the Lie algebra:
\beq
\frac{x_{a}}{\parallel
\vec{x}\parallel}\,\mapsto\,\hat{x}_{a}^{\left(N\right)}\,\,\,\,\,\,\,\,
\left[\hat{x}_{a}^{\left(N\right)},\hat{x}_{b}^{\left(N\right)}\right]=
\frac{2i\varepsilon_{abc}}{\sqrt{N^{2}-1}}
\,\hat{x}_{c}^{\left(N\right)} \ .
\eeq
The commutation rules satisfied by generators of the algebras in
the sequence $\mathcal{A}^{\left(N\right)}\left(S^{2},*\right)$
make it intuitively clear that the limit for
$N\,\rightarrow\,\infty$ of this sequence is an abelian algebra.
This is the reason why this sequence is called \emph{fuzzy
sphere}.

\section{The fuzzy disc\label{se:disc}}

In the previous section, the fuzzy sphere has been introduced via
the Weyl-Wigner formalism, which makes use of the fact that the
2-dimensional sphere is the coadjoint orbit of the \emph{compact}
group $SU\left(2\right)$. Compactness of the group brought to a
natural definition of a sequence of finite dimensional Hilbert
spaces, and a natural identification of the dimension of these
spaces as a cut-off index for a suitable expansion of a generic
function on the sphere. Properties of the sphere as an orbit of
that group made it possible to formalize every point of the sphere
as a coherent state, defined on each of those finite dimensional
spaces carrying the UIRR's. So the map between operators (finite
rank matrices) and functions on the sphere has been naturally
introduced via the Berezin procedure, and has been refined
stressing the role of the fuzzy Laplacian operator in the
definition of  fuzzy harmonics  as a basis in the set
$\mathbb{M}_{N}$. Moreover, compactness of the group played a
fundamental role in the rigorous analysis of the convergence
performed by Rieffel~\cite{rieffelGHdist}.

To approach the problem of defining a fuzzy approximation to the
disc,  it would be natural then  to analyse the possibility that
the disc were a coadjoint orbit for a Lie group. If this were the
case, it would be possible to introduce a system of coherent
states labelled by its points, and some sort of Berezin
map~\cite{berezinmap}. However this approach meets some troubles.
The group from which it is possible to define a system of coherent
states in correspondence with points of a disc is
$SU\left(1,1\right)$, but it is non compact, so its UIRR's are not
realized on finite dimensional Hilbert spaces. Hence, there  is no
intrinsic concept of a cut-off index, the would be  dimension of
the fuzzification, in this context. In~\cite{beatles} the problem
has been circumvented using coherent states for the
Heisenberg-Weyl group and the standard isomorphism in terms of
Berezin symbols between operators and functions on the plane. Then
a sequence of  projections converging to the characteristic
function of the disc has been defined. The projections identify a
sequence of finite rank matrix algebras which can be endowed with
an additional structure, a fuzzy Laplacian, identifying  the
underlying geometry. In the commutative limit $N\theta=R^2$ this
geometry is seen to converge to a disc of radius $R$.

It is well known that it is possible to define a  basis for the
space of functions on the disc in terms of  a suitable set of
Bessel functions. Since  the Weyl-Wigner map for functions on the
sphere has been introduced~(\ref{WWmapsphere}) mapping spherical
harmonics into fuzzy harmonics, we pose the question whether or
not  it is possible to define a system of \emph{fuzzy Bessel}
functions in terms of finite rank matrices, and then map again
continuum Bessel functions into them.  Since Bessel functions are
defined on the plane, in connection with  boundary values problems
for Laplacian operators, the  noncommutative plane which we are
going to revue, together with fuzzy derivations and a fuzzy
Laplacian  will turn out to be a natural setting to perform the
analysis. The final answer to the problem will be given studying
the spectral resolution of the fuzzy Laplacian.

\subsection{The noncommutative plane as a matrix algebra}

The noncommutative plane can be defined using a Weyl-Wigner map,
following again the general procedure of Berezin: so the first
step will be the definition of a set of generalised coherent
states for the Heisenberg-Weyl group, since they are labelled by
points of a plane. In this space it will be assumed a system of
coordinates of the kind $\left(x,y\right)$ for a point of
$\real^{2}$; $\theta$ will be a parameter representing an explicit
noncommutativity.

With this notation, Heisenberg-Weyl group is a manifold
$\real^{3}$, whose points are represented by a triple
$\left(x,y,\lambda\right)$, with the composition rule: \beq
\left(x,y,\lambda\right)\cdot
\left(x^{\prime},y^{\prime},\lambda^{\prime}\right)=
\left(x+x^{\prime},y+y^{\prime},\lambda+\lambda^{\prime}-
\frac{1}{\theta}\left(xy^{\prime}-x^{\prime}y\right)\right) \ .
\eeq The identity of this group is given by: \beq
id_{W}=\left(0,0,0\right)\ , \eeq and the inverse of a generic
element is: \beq
\left(x,y,\lambda\right)^{-1}=\left(-x,-y,-\lambda\right)  \ .
\eeq Considering the complex plane via $z=x+iy$ and $\bz=x-iy$ the
group law becomes: \beq
\left(z,\lambda\right)\cdot\left(z^{\prime},\lambda^{\prime}\right)=
\left(z+z^{\prime},\lambda+\lambda^{\prime}+
\frac{i}{2\theta}\left(\bz z^{\prime}-\bz^{\prime}z\right)\right)
\ . \eeq

The Hilbert space on which representing this group is again the
Fock space $\mathcal{F}$ of finite norm complex analytical
functions in the $w$ variable where the norm is obtained by the
scalar product:
\beq
\langle f|
g\rangle\,\equiv\,\int\,\frac{d^{2}w}{\pi\theta}\,e^{-\baw w
/\theta}\,\bar{f}\left(w\right)g\left(w\right)\ ,
\eeq
and an orthonormal basis is given by:
\beq
\psi_{n}\left(w\right)=\frac{w^{n}}{\sqrt{\theta^{n}n!}}
\ . \eeq The unitary representation of the Heisenberg-Weyl group
is given by operators
$\To\left(z,\lambda\right)\,\in\,Op\left(\mathcal{F}\right)$: \beq
\left(\To f\right)\left(w\right)= e^{i\lambda}e^{-\baz z
/2\theta}e^{z w/\theta} f\left(w-\baz\right) \label{unirephw} \ .
\eeq

Chosen the first basis element $\psi_{0}\left(w\right)$ as the
fiducial state, the procedure already outlined gives a coherent
state for each point of the complex plane. In the realization of
the Fock space as a space of complex analytical functions, a
coherent state is then: \beq |
z\rangle\,\,\,\rightarrow\,\,\,\psi_{z}\left(w\right)= e^{-\baz
z/2\theta}e^{zw/\theta} \eeq and, given $\psi_{n}$ an element of
the basis already considered: \beq | z\ra =\sum_{n=0}^{\infty}
e^{-\baz z/2\theta}\frac{z^{n}}{\sqrt{n!\theta^{n}}}|\psi_{n}\ra \
. \eeq Such coherent states are non orthogonal, and overcomplete:
\beqa \la z|
z^{\prime}\ra&=&e^{-\left(\left|z\right|^{2}+\left|z^{\prime}\right|^{2}-2\baz
z^{\prime}\right)/\theta} \ ,\nonumber \\
\id&=&\int\frac{d^{2}z}{\pi\theta}\,| z \rangle\langle
z|\label{HWcomplete} \ . \eeqa

On this Hilbert space $\mathcal{F}$, it is possible to introduce a
pair of creation-annihilation operators: \beqa
\left(\hat{a}f\right)\left(w\right)&=&\theta\frac{df}{dw}
\nonumber
\\
\left(\hat{a}^{\dagger}f\right)\left(w\right)&=&wf\left(w\right)
 \ ,
\eeqa
such that:
\be
\left[\hat{a},\hat{a}^{\dagger}\right]=\theta\,\id,
\label{commaadagger}
\ee
and:
\be
\hat a|z\rangle=z|z\rangle \ ,
\ee
so that we can recognize the usual coherent states encountered in
quantum mechanics textbooks. In the chosen orthonormal basis, one
has:
\beqa \hat{a}| \psi_{n}\rangle&=&\sqrt{n\theta}| \psi_{n-1}\rangle
\nonumber
\\
\hat{a}^{\dagger}|\psi_{n}\rangle&=&\sqrt{\left(n+1\right)\theta}|
\psi_{n+1}\rangle \ ,
\eeqa
and therefore one recognizes the basis as the basis of
eigenvectors of the number operator $\hat{N}=\hat{a}^{\dagger}\hat{a}$.
Note however the unconventional presence $\theta$ in the
commutation relation~\eqn{commaadagger}.

These relations can be extended. A Berezin symbol can be
associated to an operator in the Fock space:
\be
f\left(\baz,z\right)=\langle z| \fo| z\rangle\ . \ee This can be
seen as a Wigner map. It can be inverted:
\be
\fo=\int\,\frac{d^{2}\xi}{\pi\theta}\,
\int\,\frac{d^{2}z}{\pi\theta}\,f\left(z,\bar{z}\right)\,
e^{-\left(\bar{z}\xi-\bar{\xi}z\right)/\theta}\,e^{\xi\hat{a}^{\dagger}/\theta}\,
e^{-\bar{\xi}\hat{a}/\theta}. \label{discweylmap}\ee

This quantization map for functions on a plane can be given an
interesting form. To start with, we can restrict to functions which
can be written as Taylor series in $\baz,z$: \beq
f\left(\baz,z\right)=\sum_{m,n=0}^{\infty}\,f_{mn}^{Tay}\baz^{m}z^{n}
\ . \eeq An easy calculation shows that this $f$ is the symbol of
the operator: \beq
\fo=\sum_{m,n=0}^{\infty}\,f_{mn}^{Tay}\ao^{\dagger m}\ao^{n}
\label{taylorexp}\ . \eeq More generally we can consider operators
written in a density matrix notation: 
\beq
\fo=\sum_{m,n=0}^{\infty}\,f_{mn}|\psi_{m}\ra\la\psi_{n}| \ .
\eeq 
The Berezin symbol of this operator is the function: 
\beq
f\left(\baz,z\right)=e^{-\left|z\right|^{2}/\theta}
\sum_{m,n=0}^{\infty}\,f_{mn}\frac{\baz^{m}z^{n}}{\sqrt{m!n!\theta^{m+n}}}
\label{dmsymbol}
 \ , 
\eeq 
where the relation between the Taylor coefficients $f_{mn}^{Tay}$
and  the $f_{mn}$ is  \beq
f_{lk}=\sum_{q=0}^{\min\left(l,k\right)}\,f_{l-q\,k-q}^{Tay}\,
\frac{\sqrt{k!l!\theta^{l+k}}}{q!\theta^{q}} \ , \eeq while the
inverse relation is given by: 
\beq
f_{mn}^{Tay}=\sum_{p=0}^{\min\left(m,n\right)}\,
\frac{\left(-1\right)^{p}}{p!\sqrt{\left(m-p\right)!\left(n-p\right)!\theta^{m+n}}}
\,f_{m-p\,n-p}\ .
\eeq 
Equation~(\ref{taylorexp}) shows that the
quantization of a monomial in the variables $z,\baz$ is an
operator in $\ao,\aod$, formally a monomial in these two
noncommuting variables, with all terms in $\aod$ acting at the
left side with respect to terms in $\ao$.  This ordering is
related to the presence, in the quantization
map~(\ref{discweylmap}), of a specific term that, in the standard
terminology used for the Weyl-Wigner formalism, is called
weight~\cite{wignerreview}. In fact, restoring real variables
~\eqn{discweylmap} can be written as:
\beq
\hat{f}=\int\,\frac{dadb}{2\pi\theta}\,
\int\,\frac{dxdy}{2\pi\theta}\,f\left(x,y\right)\,
e^{-\frac{i\left(bx-ay\right)}\theta}\,
e^{\frac{u\hat{a}^{\dagger}}\theta}\,e^{-\frac{\bar{u}\hat{a}}\theta}
=\int\,\frac{dadb}{2\pi\theta}\,\tilde{f}\left(b,a\right)\,
e^{\frac{\left(a^{2}+b^{2}\right)}{8\theta}}\,
e^{\frac{\left(u\aod-\bar{u}\ao\right)}\theta}
 \label{weightdisc}
\ . 
\eeq 
with  $u=\left(a+ib\right)/2$.   The function $\tilde{f}$
is the \emph{symplectic} Fourier transform of the function $f$.
Compared~\eqn{weightdisc} with the standard Weyl
map~\cite{wignerreview,zampthesis}, the difference is the factor
$e^{\left(a^{2}+b^{2}\right)/8\theta}$.

The invertibility of the Weyl map (on a suitable domain of
functions on the plane) enables to define a noncommutative product
in the space of functions,  known as Voros
product~\cite{vorosproduct,Aletal}, a variant of the more popular
Gr\"onewold-Moyal product~\cite{Gronewold,Moyal}:
\beq \left(f*g\right)\left(\baz,z\right)=\la z|\fo\go| z\ra
\label{vorospr}\ . \eeq It is a non local product: \beq
\left(f*g\right)\left(\baz,z\right)= e^{-\baz
z/\theta}\int\,\frac{d^{2}\xi}{\pi\theta}\,
f\left(\baz,\xi\right)g\left(\bar{\xi},z\right)
\,e^{-\bar{\xi}\xi/\theta}e^{\bar{\xi}z/\theta}e^{\baz\xi/\theta}
\ . 
\eeq 
Its asymptotic expansion, on a suitable domain, acquires
the form: 
\beq 
\left(f*g\right)\left(\baz,z\right)=f\,
e^{\theta\overleftarrow{\del}_{\baz}\overrightarrow{\del}_{z}}\,g,
\eeq 
and makes it clear that it is a deformation, in $\theta$, of
the pointwise commutative product. Since it is the translation, in
the space of functions, of the product in the space of operators,
if symbols are expressed in the form~(\ref{dmsymbol}), then the
product acquires a matrix form: 
\beq
\left(f*g\right)_{mn}=\sum_{k=0}^{\infty}f_{mk}g_{kn}
\label{matrixprod}\ . 
\eeq

The space of functions on the plane, with the standard definition
of sum, and the product given by the Voros product
(\ref{vorospr}), is a nonabelian algebra, a noncommutative plane.
This algebra
$\mathcal{A}_{\theta}=\left(\mathcal{F}\left(\real^{2}\right),*\right)$
is isomorphic to an algebra of operators, or, as equation
(\ref{matrixprod}) suggests, to an algebra of infinite dimensional
matrices. A more detailed mathematical analysis of the Moyal algebra, i.e. of the quantum plane,
is given in \cite{marseticos} and references therein.

\subsection{A sequence of non abelian algebras}

A fuzzy space has been presented as a sequence of finite rank
matrix algebras converging to an algebra of functions. 
The meaning of this convergence is formalised as compact 
quantum metric spaces. In the case of the fuzzy sphere the rank
of the matrices involved is the dimension of the Hilbert space on
which UIRR's of the group $SU\left(2\right)$ are realised. In the
approach sketched in the last section there is no natural
definition of a set of finite dimensional matrix algebras. This is
a consequence of the fact that the noncommutative plane has been
realised via a Berezin quantization based on coherent states
originated by a group, the Heisenberg-Weyl, which is noncompact.

In this context the strategy to obtain finite dimensional matrix
algebras is different. $\mathcal{A}_{\theta}$ can be considered,
once a basis in the Hilbert space has been chosen, as a matrix
algebra made up by  formally infinite dimensional matrices. One
can define a set of finite dimensional matrix algebras simply
truncating $\mathcal{A}_{\theta}$. The notion of truncation is
formalised via the introduction of a set of projectors:
\be
\hat{ P}_{\theta}^{(N)}= \sum_{n=0}^{N} |\psi_n\rangle\langle\psi_n|
\ee
in the space of operators. Their symbols are projectors in the
algebra $\mathcal{A}_{\theta}$ of the noncommutative plane, in the
sense that they are idempotent functions of order $2$ with respect
to the Voros product (here $z=re^{i\varphi}$): \beqa
P_{\theta}^{\left(N\right)}\left(r,\varphi\right)&=&
\sum_{n=0}^{N}\la z|\psi_{n}\ra\la\psi_{n}| z\ra=e^{-r^{2}/\theta}
\sum_{n=0}^{N}\,\frac{r^{2n}}{n!\theta^{n}} \nonumber \\
P_{\theta}^{\left(N\right)}*P_{\theta}^{\left(N\right)}&=&
P_{\theta}^{\left(N\right)} \ . \eeqa This finite sum can be
performed yielding a rotationally symmetric function: \beq
P_{\theta}^{\left(N\right)}\left(r,\varphi\right)\,=
\,\frac{\Gamma\left(N+1,r^{2}/\theta\right)}{\Gamma\left(N+1\right)}
\ . \eeq in terms of the ratio of an incomplete gamma function by
a gamma function~\cite{prudnikov}. If $\theta$ is kept fixed, and
nonzero, in the limit for $N\,\rightarrow\,\infty$ the symbol
$P_{\theta}^{\left(N\right)}\left(r,\varphi\right)$ converges,
pointwise, to the constant function
$P_{\theta}^{\left(N\right)}\left(r,\varphi\right)=1$, which can
be formally considered as the symbol of the identity operator: in
this limit one recovers the whole noncommutative plane.

This situation changes if the limit for $N\,\rightarrow\,\infty$
is performed keeping the product $N\theta$ equals to a constant,
say $R^{2}$. In a pointwise convergence, chosen $R^{2}=1$: \beq
P^{\left(N\right)}_{\theta}\,\,\rightarrow\,\,\,\left[\begin{array}{cc}
1 & r<1 \\ 1/2 & r=1 \\ 0 & r> 1
\end{array}\right]=Id\left(r\right) \ .
\eeq
This sequence of projectors converges to a step function in the
radial coordinate $r$, the characteristic function of a disc on
the plane. Thus a sequence of subalgebras
$\mathcal{A}_{\theta}^{\left(N\right)}$ can be defined by: \beq
\mathcal{A}_{\theta}^{\left(N\right)}=P_{\theta}^{\left(N\right)}*
\mathcal{A}_{\theta} *P_{\theta}^{\left(N\right)} \ . \eeq As it
has been said, the full algebra $\mathcal{A}_{\theta}$ is
isomorphic to an algebra of operators. What the previous relation
says is that $\mathcal{A}_{\theta}^{\left(N\right)}$ is isomorphic
to $\mathbb{M}_{N+1}$, the algebra of $\left(N+1\right)$ rank
matrices: the important thing is that this isomorphism (this
truncation) is obtained via a specific choice of a basis in the
Fock space $\mathcal{F}$ on which coherent states for the
Heisenberg-Weyl group are realised. The  effect of this projection
on a generic function is: \beq
\Pi^{\left(N\right)}_{\theta}\left(f\right)=
f^{\left(N\right)}_{\theta}=P^{\left(N\right)}
_{\theta}*f*P^{\left(N\right)}_{\theta}=
e^{-\left|z\right|^{2}/\theta}
\sum_{m,n=0}^{N}f_{mn}\frac{\baz^{m}z^{n}}{\sqrt{m!n!\theta^{m+n}}}\
.
\eeq On every subalgebra $\mathcal{A}^{\left(N\right)}_{\theta}$,
the symbol $P^{\left(N\right)}_{\theta}\left(r,\varphi\right)$ is
then the identity, because it is the symbol of the projector
$\Po^{\left(N\right)}=\sum_{n=0}^{N}|\psi_{n}\ra\la\psi_{n}|$,
which is the identity operator in
$\mathcal{A}_{\theta}^{\left(N\right)}$, or, equivalently, the
identity matrix in every $\mathbb{M}_{N+1}$.

Note that the rotation group on the plane, $SO\left(2\right)$,
acts in a natural way on these subalgebras. Its generator is the
truncated number operator
$\No^{(N)}=\sum_{n=0}^{N}\,n\theta|\psi_{n}\ra\la\psi_{n}|$.
Cutting at a finite $N$ the expansion provides an infrared cutoff.
This cutoff is ``fuzzy" in the sense that functions in the
subalgebra are still defined outside the disc of radius $R$, but
are exponentially damped.  In general  functions are close to
their projected version $f^{\left(N\right)}_{\theta}$ if they are
mostly supported on a disc of radius  $R=\sqrt{N\theta}$,
otherwise they are  exponentially cut, provided they  do not
present oscillations of too small wavelength (compared to
$\theta$). In this case the projected function becomes very large
on the boundary of the disc. More details and examples are
in~\cite{beatles,balfest}.

\subsection{Fuzzy derivatives and fuzzy Laplacian}

So far the Weyl-Wigner formalism, and the projection procedure,
have provided a way to associate to functions on the plane a
sequence of finite dimension
$\left(N+1\right)\times\left(N+1\right)$ matrices. An appropriate
choice of $\theta$, the noncommutativity parameter introduced by
the quantization map, and $N$, showed that it is possible to
obtain a good approximation of a certain class of functions
supported on a disc. The next step is the analysis of the geometry
these algebras can formalize. To pursue this task, we need
derivations and a Laplacian. The starting point to define the
matrix equivalent of the derivations is:
 \bea
\del_{z}f\,&=&\,\frac{1}{\theta}\langle z|
\left[\hat{f},\hat{a}^{\dagger}\right]| z\rangle \ , \nonumber \\
\del_{\bar{z}}f\,&=&\,\frac{1}{\theta}\langle z|
\left[\hat{a},\hat{f}\right]| z\rangle \ . \label{exactdersym}
\eea
This relation is exact in the full algebra $\mathcal{A}_{\theta}$.
Given an operator $\fo$, the derivatives of the symbol
$f\left(\baz,z\right)$ are related to the symbol of the commutator
of $\fo$ with the creation and annihilation operators. A fuzzified
version, namely a truncated version, of these operations, can be
defined as\footnote{This a slightly modified version of the
derivations in~\cite{beatles,balfest}}:
\bea
\del_{z}f^{\left(N\right)}_{\theta}&\equiv
&\frac{1}{\theta} \langle z|\hat{P}^{\left(N\right)}_{\theta}
\left[\hat{P}^{\left(N\right)}_{\theta}\hat{f}\hat{P}^{\left(N\right)}_{\theta},
\hat{a}^{\dagger}\right]
\hat{P}^{\left(N\right)}_{\theta}| z\rangle \nonumber \\
\del_{\bar{z}}f^{\left(N\right)}_{\theta}& \equiv
&-\frac{1}{\theta} \langle z|\hat{P}^{\left(N\right)}_{\theta}
\left[\hat{P}^{\left(N\right)}_{\theta}\hat{f}\hat{P}^{\left(N\right)}_{\theta},
\hat{a}\right] \hat{P}^{\left(N\right)}_{\theta}| z\rangle\ .
\label{fuzzder}
\eea
It is important to note that this is really a derivation on each
$\mathcal{A}^{\left(N\right)}_{\theta}$, that is  a linear
operation from $\mathcal{A}^{\left(N\right)}_{\theta}$ to itself,
satisfying the Leibnitz rule. The way this is achieved
in~\eqn{fuzzder} goes through the following steps. We first
associate to $f^{\left(N\right)}_{\theta}$ a finite rank matrix
$\fo^{\left(N\right)}_{\theta}$ which is indeed regarded as
embedded in an infinite dimensional one, with all zeroes but
$(N+1) \times (N+1)$ elements in the upper left corner:
$\fo^{\left(N\right)}_{\theta}=\hat{P}^{\left(N\right)}_{\theta}\fo
\hat{P}^{\left(N\right)}_{\theta} $.  We then perform the
commutator with $\hat{a},~\hat{a}^{\dagger}$ realised as infinite
dimensional matrices. This operation yields an infinite
dimensional matrix with all zeroes but $(N+2) \times (N+2)$
elements in the upper left corner, so that we have to project back
to $\mathbb{M}_{N+1}$ to obtain a properly defined derivation.
Finally, we consider the symbol.

As an example let us  calculate the derivatives of the fuzzified
coordinate functions $z$ and $\bar{z}$. From the definition, $z$,
$\bar{z}$ are respectively the   symbol of the annihilation and
creation operators, which  can be written as:
\bea
\hat{a}\,&=&\,\sum_{s=0}^{\infty}\,\sqrt{\left(s+1\right)\theta}|
\psi_{s}\rangle\langle\psi_{s+1}| \nonumber
\\
\hat{a}^{\dagger}\,&=&\,\sum_{k=0}^{\infty}\,\sqrt{\left(k+1\right)\theta}|
\psi_{k+1}\rangle\langle\psi_{k}|.
\eea 
Projection into $\mathbb{M}_{N+1}$  gives: 
\beq
\hat{a}^{\left(N\right)}_{\theta}=\sum_{s=0}^{N-1}\,
\sqrt{\left(s+1\right)\theta}|\psi_{s}\rangle\langle\psi_{s+1}| \
.
\eeq 
According to \eqn{fuzzder},  to perform the derivative with
respect to $z$, we have to  consider: 
\bea
\left[\hat{a}^{\left(N\right)}_{\theta},\hat{a}^{\dagger}\right]&=&
\theta\left[|\psi_{0}\rangle\langle\psi_{0}|+\sum_{s=1}^{N-1}|
\psi_{s}\rangle\langle\psi_{s}|-N|\psi_{N}\rangle\langle\psi_{N}|\right]
\nonumber
\\
&=&\theta\left[\sum_{s=0}^{N-1}|\psi_{s}\rangle\langle\psi_{s}|-
N|\psi_{N}\rangle\langle\psi_{N}|\right] \nonumber
\\
&=&\theta\left[\sum_{s=0}^{N}|\psi_{s}\rangle\langle\psi_{s}|-\left(1+N\right)
|\psi_{N}\rangle\langle\psi_{N}|\right]. 
\eea 
The first term of
the sum is the projector onto the first $N+1$ basis elements,
whose symbol is the identity in
$\mathcal{A}^{\left(N\right)}_{\theta}$. It is worth noting that
this commutator has no terms 'outside' the space we are
considering, namely there are no components on density matrices of
order greater than $N+1$: this means that in this case there is no
need to project it on $ \mathbb{M}_{N+1}$. The symbol yields then:
\beq \del_{z}\left(z^{\left(N\right)}_{\theta}\right)
=e^{-Nr^{2}}\left[\sum_{s=0}^{N}\,\frac{r^{2s}N^{s}}{s!}\,-\,
\frac{N+1}{N!}r^{2N}N^{N}\right] \ .  \eeq In the limit of
$N\rightarrow\infty$ the first term is the characteristic function
for the disc, while the second converges to a factor
$\pi\delta\left(r-1\right)$. This factor is a radial $\delta$
selecting the value  $r=1$ with respect to the Lebesgue measure on
the plane: \beq
\lim_{N\rightarrow\infty}\del_{z}\left(z^{\left(N\right)}_{\theta}\right)
=Id\left(r\right)-\pi\delta\left(r-1\right)\ . \eeq To calculate
the derivative of $\bar{z}$ with respect to $z$ one needs to
consider:
$$\hat{a}^{\dagger\left(N\right)}_{\theta}=
\sum_{k=0}^{N-1}\,\sqrt{\left(k+1\right)\theta}
|\psi_{k+1}\rangle\langle\psi_{k}|$$ and the commutator:
\beq
\left[\hat{a}^{\dagger\left(N\right)}_{\theta},\hat{a}^{\dagger}\right]
=-\theta\sqrt{N\left(N+1\right)}|\psi_{N+1}\rangle\langle\psi_{N}
| \ . \eeq Unlike the previous example this operator must be
projected back to the algebra $\mathbb{M}_{N+1}$, and upon
considering the symbol we  finally have: \beq
\del_{\bar{z}}\left(z^{\left(N\right)}_{\theta}\right)=0\ . \eeq

Let us come to  the definition of the Laplacian operator. In the
spirit of noncommutative geometry it is this additional structure
which carries the information about the geometry of the space
underlying $\mathcal{A}_{\theta}$. In facts, we can say we have
succeeded in fuzzifying  the disc only if we have been able to
define a fuzzy Laplacian  whose spectrum approaches that of the
ordinary Laplacian on the disc when $N\rightarrow \infty$.

Starting from the exact expressions:
\be
\nabla^2\,f(\bar z,z)=4\del_{\bar{z}}\del_{z}f
=\frac{4}{\theta^{2}}\la z|\left[\hat{a},
\left[\hat{f},\hat{a}^{\dagger}\right]\right]| z\ra
\ee
it is possible to define, in each
$\mathcal{A}^{\left(N\right)}_{\theta}$: \beq
\nabla^2_{\left(N\right)}\,\fo^{\left(N\right)}_{\theta}
\equiv\frac{4}{\theta^{2}}\hat{P}^{\left(N\right)}_{\theta}
\left[\hat{a},\left[\hat{P}^{\left(N\right)}_{\theta}
\hat{f}\hat{P}^{\left(N\right)}_{\theta},\hat{a}^{\dagger}\right]\right]
\hat{P}^{\left(N\right)}_{\theta} \label{lapl} \ . \eeq

The image of the element of the truncated algebra:
$$\hat{f}^{\left(N\right)}_{\theta} =
\sum_{a,b=0}^{N}\,f_{ab}|\psi_{a} \rangle \langle\psi_{b}|$$ is
then: 
\bea
\nabla^{2}_{\left(N\right)}\,f^{\left(N\right)}_{\theta}&
=&\,4N\,\left[\sum_{s=0}^{N-1}\,\sum_{b=0}^{N-1}\,f_{s+1,b+1}\sqrt{\left(s+1\right)
\left(b+1\right)}|\psi_{s}\rangle\langle\psi_{b}|\right.+ \nonumber
\\
&&-\sum_{s=0}^{N}\,\sum_{b=0}^{N}\,f_{sb}\left(s+1\right)
|\psi_{s}\rangle\langle\psi_{b}|\,-\,
\sum_{s=0}^{N-1}\,f_{0,s+1}\left(s+1\right)
|\psi_{0}\rangle\langle\psi_{s+1}|+ \nonumber
\\
&&+\sum_{s=0}^{N-1}\,\sum_{b=0}^{N-1}\,f_{sb}
\sqrt{\left(s+1\right)\left(b+1\right)}|\psi_{s+1}\rangle
\langle\psi_{b+1}|+\nonumber
\\
&& -\left.\sum_{s=0}^{N-1}\,\sum_{b=0}^{N-1}\,f_{s+1,b+1}\,
\left(b+1\right)|\psi_{s+1}\rangle\langle\psi_{b+1} |\,\right].
\label{explapl}
\eea
The eigenvalues of this Laplacian can be numerically calculated.
They are seen to converge to the spectrum of the continuum one for functions
on a disc, with boundary conditions on the edge of the disc of
Dirichlet homogeneous kind~\cite{tichonov}. All eigenvalues are
negative, and their modules $\lambda$ solve the implicit
equation:
\beq
J_{n}\left(\sqrt{\lambda}\right)=0\ ,
\eeq
where $n$ is the order of the Bessel functions. In particular, those
related to $J_{0}$ are simply degenerate, the others are doubly
degenerate: so there is a sequence of eigenvalues  labelled by
$\lambda_{n,k}$ where $n$ is the order of the Bessel function and
$k$ indicates that it is the $k^{th}$ zero of the function. The
eigenfunctions of this continuum operator are:
\beq
\Phi_{n,k}=e^{in\varphi}
\left(\frac{\sqrt{\lambda_{\left|n\right|,k}}r}{2}\right)^{|n|}
\sum_{s=0}^{\infty}
\frac{\left(-\lambda_{|n|,k}\right)^{s}}{s!\left(|n|+s\right)!}
\left(\frac{r}{2}\right)^{2s}=
e^{in\varphi}J_{|n|}\left(\sqrt{\lambda_{\left|n\right|,k}}r\right)
\ . \label{bessels}
\eeq
with  $n$ integer number and $|n|$  its absolute value. This is a
way to write the eigenfunctions in a compact form, taking into
account the degeneracy of eigenvalues for $|n|\geq 1$.

\begin{figure}[htbp]
\epsfxsize=2.3 in \centerline{\epsfxsize=2.2
in\epsffile{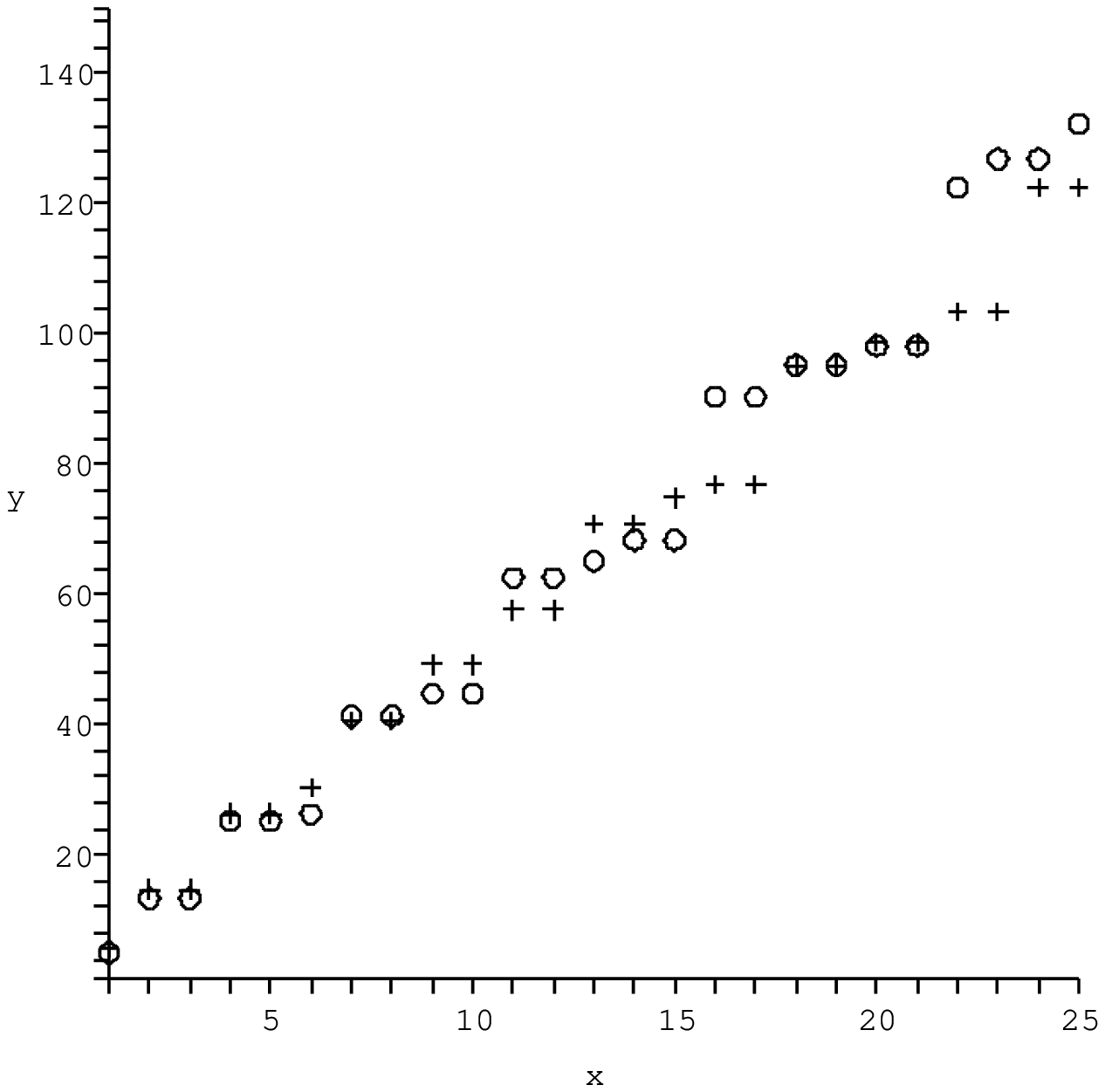}\epsfxsize=2.2
in\epsffile{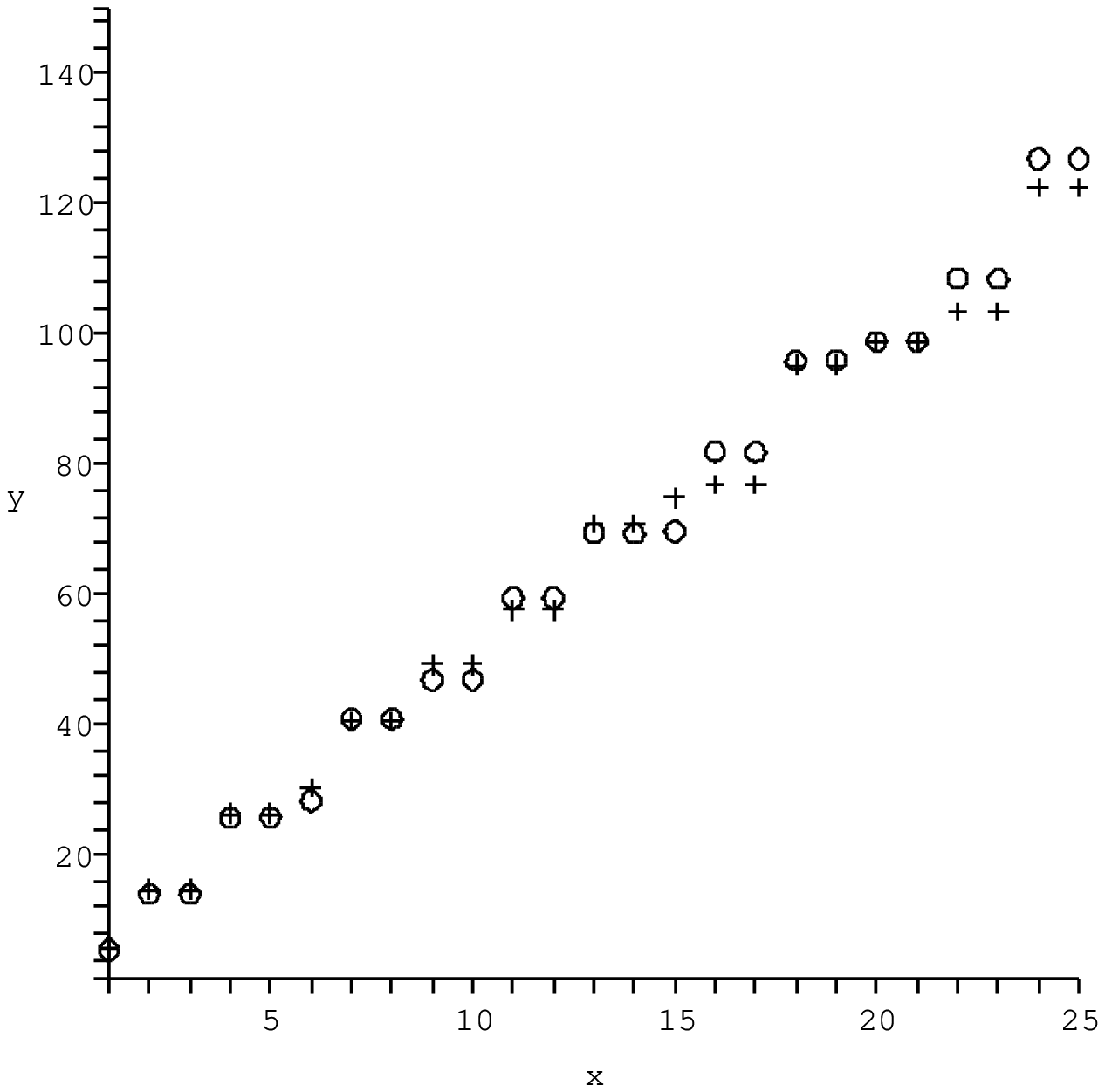}\epsfxsize=2.2
in\epsffile{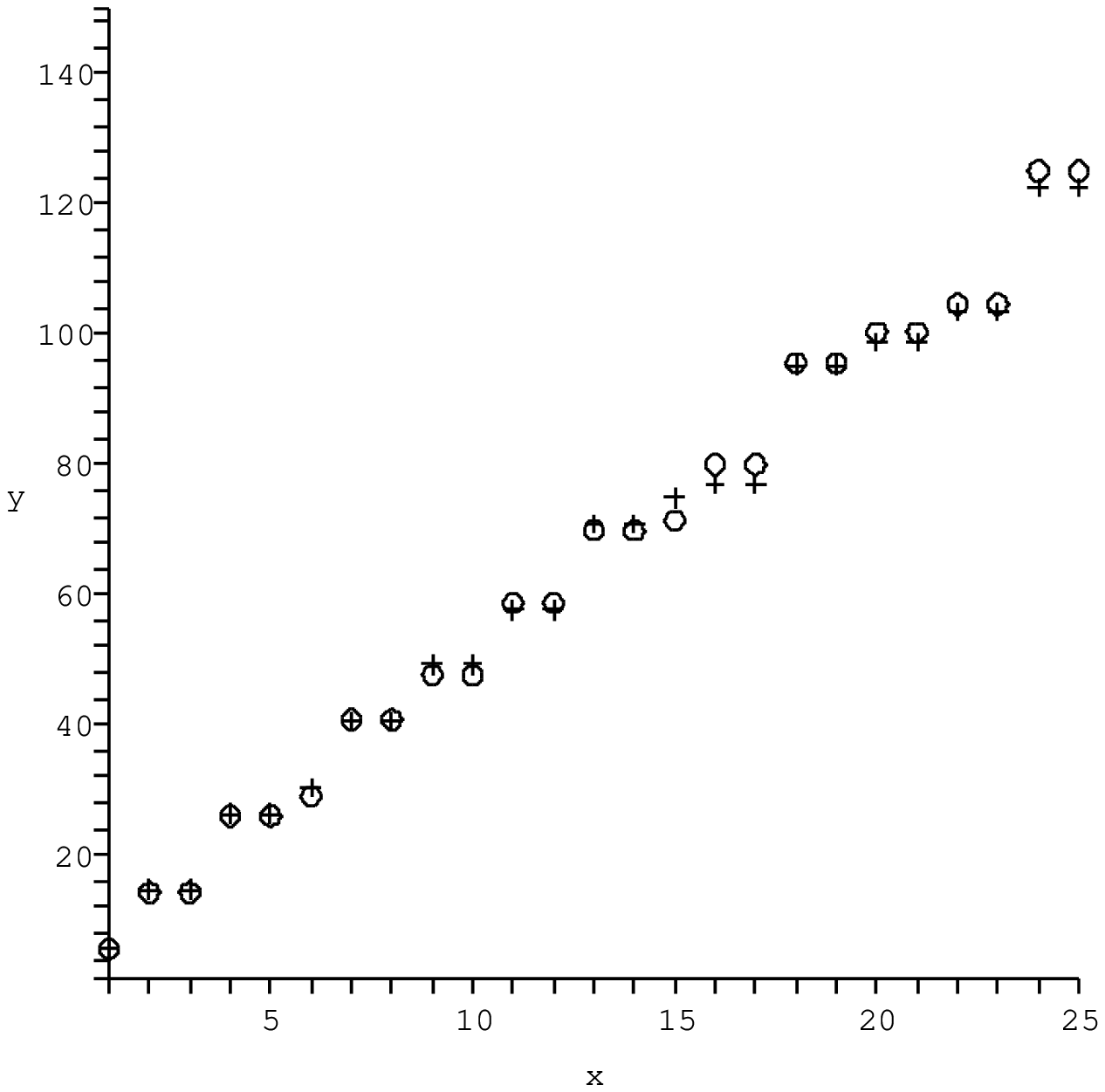}} \caption{\baselineskip=12pt {\it
Comparison of the first eigenvalues of the fuzzy Laplacian
(circles) with those of the continuum Laplacian (crosses) on the
domain of functions with Dirichlet homogeneous boundary
conditions. The orders of truncation are $N=10,20,30$.}}
\bigskip
\label{fuzzydrum}
\end{figure}

The spectrum of the fuzzy Laplacian is in good agreement with the
spectrum of the continuum case, even for low values $N$ of the
dimension of truncation, as can be seen in figure~\ref{fuzzydrum}.
It is interesting to note  that with this definition of Laplacian
(\ref{lapl}) we correctly reproduce  the  pattern of non
degenerate and double degenerate eigenvalues. The difference with
the case of the spectrum of the fuzzy Laplacian for the fuzzy
sphere is that now the "fuzzy spectrum" is both  a cut-off and an
approximation of the continuum spectrum. It is a cut-off because,
of course, it is a finite rank operator. It is an approximation
because it  has been defined using a formalism whose building
blocks are related to a noncompact group, namely the
Heisenberg-Weyl, whose generators have no finite  dimensional
realization.

\section{Fuzzy Bessel functions\label{se:fuzzybessel}}
\setcounter{equation}{0}

In this section we introduce a basis on the fuzzy disc. As we have
already noticed in the introduction, we cannot use representation
theory, which in the case of the fuzzy sphere yields the fuzzy
harmonics, because there is no compact group whose homogeneous
space is the disc. But we have a well defined Laplacian for each
$\mathcal{A}_\theta^{\left(N\right)}$. Therefore we may answer the
problem looking at the eigenfunctions of this fuzzy Laplacian.

\subsection{Spectral properties of the Laplacian \label{specprop}}
The fuzzy Laplacians~(\ref{lapl}) are a family of operators
mapping each algebra $\mathcal{A}^{\left(N\right)}_{\theta}$ into
itself with the explicit action~(\ref{explapl}). Considering
$\mathcal{A}^{\left(N\right)}_{\theta}$ as a
$\left(N+1\right)^{2}$ dimensional vector space, a density matrix
$|\psi_{a}\rangle\langle\psi_{b}|$ is a basis element for it, with
$a,b\,\in\,\left(0,\ldots,N\right)$. Equation~(\ref{explapl}) can
be specified:
\bea
\nabla^{2}_{\left(N\right)}\left(|\psi_{a}\rangle\langle\psi_{b}|\right)&=&
4N\,\Po^{\left(N\right)}_{\theta}\left(\sqrt{ab}\,
|\psi_{a-1}\rangle\langle\psi_{b-1}|\right.+
\\&&\left.-\left(a+b+1\right)\,
|\psi_{a}\rangle\langle\psi_{b}|\,+\,\sqrt{\left(a+1\right)\left(b+1\right)}
\,|\psi_{a+1}\rangle\langle\psi_{b+1}|\right)
\Po^{\left(N\right)}_{\theta}\ . \nonumber\label{basislapl}
\eea
The action of the projector operator is to cut away some of the
components of the right side image for $a=0,N$ and for $b=0,N$. It
is  important to note that, chosen an integer $n=a-b$, the fuzzy
Laplacian maps basis elements with the same $n$ onto themselves,
as $n$ ranges through $\left(-N,\ldots,+N\right)$. In the matrix
representation of $\mathcal{A}^{\left(N\right)}_{\theta}$, the
number $n$ fixes the diagonal to which the element
$|\psi_{a}\rangle\langle\psi_{b}|$ belongs. The
equation~(\ref{basislapl}) then shows that the fuzzy Laplacian
maps each diagonal subspace of $\mathbb{M}_{N+1}$ into itself.
This property simplifies the study of the spectral properties of
the fuzzy Laplacian. The analysis of these spectral properties
will be performed restricting the operator to each of its
stability subspaces.

To understand the meaning of this integer $n$, go back to the
eigenfunctions of the continuum Laplacian on a disc with Dirichlet
homogeneous boundary conditions~(\ref{bessels}). Eigenfunctions
$\Phi_{n,k}$ are defined on the real plane, so they can be mapped
into operators on the whole Hilbert space via the Weyl map
introduced before~(\ref{discweylmap}). The continuum
eigenfunctions~(\ref{bessels})
with a fixed nonnegative $n$, such that $0\leq n\leq N$ can be
mapped into an operator:
\beq
\hat{\Phi}_{n,k}=\left(\frac{\lambda_{n,k}}{4}\right)^{n/2}\,\sum_{s=0}^{\infty}
\,\left(-\frac{\lambda_{n,k}}{4}\right)^{s}\,\frac{1}{s!\left(s+n\right)!}\,
\ao^{\dagger s}\,\ao^{s+n} \ . \eeq In the density matrix
notation, it acquires the form: \beq \hat{\Phi}_{n,k}=
\left(\frac{\lambda_{n,k}}{4}\right)^{n/2}\, \sum_{j=n}^{\infty}\,
\sum_{s=0}^{j-n}
\,\left(-\frac{\theta\,\lambda_{n,k}}{4}\right)^{s}\,
\frac{\theta^{n/2}}{s!\left(s+n\right)!}\,
\,\frac{\sqrt{j!\left(j-n\right)!}}{\left(j-s-n\right)!}\,
|\psi_{j-n}\rangle\langle\psi_{j}| \ . \eeq This element can be
fuzzified, so that now the parameter $\theta$ is constrained as $\theta N=1$: \beq \hat{\Phi}_{n,k}^{\left(N\right)}=
\left(\frac{\lambda_{n,k}}{4N}\right)^{n/2}\,
\sum_{a=0}^{N-n}\,\sqrt{a!\left(a+n\right)!}\,\left[
\sum_{s=0}^{a} \,\left(-\frac{\lambda_{n,k}}{4N}\right)^{s}\,
\frac{1}{s!\left(s+n\right)!\left(a-s\right)!}\right]\,
|\psi_{a}\rangle\langle\psi_{a+n}| \label{Phinfuzz}\ . \eeq
Eigenfunctions with negative $n$ are just the complex conjugate of
those with positive $n$:
$\left(\Phi_{-n,k}\right)^{*}=\Phi_{n,k}$. The fuzzification
procedure gives a matrix which can be obtained by
$\hat{\Phi}_{n,k}^{\left(N\right)}$ by an hermitian conjugation:
$\hat{\Phi}_{-n,k}^{\left(N\right)}=\hat{\Phi}_{n,k}^{\left(N\right)\dagger}$.
This analysis clarifies that, in the fuzzy approximation,
continuum eigenfunctions are represented by matrices belonging to
the $n^{th}$ diagonal subspace of $\mathbb{M}_{N+1}$. Moreover, in
this approach, admissible $n$ are constrained by the dimension of
the fuzzification: $n\leq N$. In all these fuzzy elements,
$\lambda_{n,k}$ (as well as $\lambda_{0,k}$) can be seen no longer
as eigenvalues of the Laplacian operator on the disc with
Dirichlet homogeneous boundary conditions, but, at this stage, simply as
parameters.

We can now perform a complete spectral analysis of the fuzzy
Laplacian. This will also show the exact meaning of the
$\lambda_{n,k}$ parameters. Consider first the $n=0$ subspace,
whose dimension is $N+1$, and consider the elements
$\hat{\Phi}_{0,k}^{\left(N\right)}$ with $k=1,\ldots\,N+1$. Act on
these elements with the fuzzy Laplacian. From
equation~(\ref{basislapl}) we obtain:
\beq
\nabla^{2}_{\left(N\right)}\left(|\psi_{a}\rangle\langle\psi_{a}|\right)=
4N\,\Po^{\left(N\right)}_{\theta}
\left(a\,|\psi_{a-1}\rangle\langle\psi_{a-1}|\,-\,\left(2a+1\right)\,
|\psi_{a}\rangle\langle\psi_{a}|\,+\,\left(a+1\right)
\,|\psi_{a+1}\rangle\langle\psi_{a+1}|\right)
\Po^{\left(N\right)}_{\theta}\label{diagobasislapl}
\eeq
one can reproject the image
$\nabla^{2}_{\left(N\right)}\left(\hat{\Phi}_{n,k}^{\left(N\right)}\right)$
on the basis density matrices for the specific subspace. On the
first basis element, equation(\ref{diagobasislapl}) becomes: \beq
\langle\psi_{0}|\
\nabla^{2}_{\left(N\right)}\left(|\psi_{a}\rangle\langle
\psi_{a}|\right)|\psi_{0}\rangle=
4N(a\delta_{0,a-1}-\left(2a+1\right)\delta_{0,a}) \ . \eeq
Considering the explicit form of~(\ref{Phinfuzz})  we obtain, for $n=0$:
\beq
\hat{\Phi}^{\left(N\right)}_{0,k}=\sum_{a=0}^{N}\,
\phi_{0,k}^{\left(a,N\right)}\,|\psi_{a}\rangle\langle\psi_{a}|;\eeq where: \beq
 \phi_{0,k}^{\left(a,N\right)}= \left[\sum_{s=0}^{a}\,
\left(-\frac{\lambda_{0,k}}{4N}\right)^{s}
\frac{1}{s!}\left(\begin{array}{c} a \\ s
\end{array}\right)\right] \ ,
\eeq
and upon applying the Laplacian:
\beqa
\langle\psi_{0}|\ \nabla^{2}_{\left(N\right)}
\left(\hat{\Phi}_{0,k}^{\left(N\right)}\right)|\psi_{0}\rangle
&=&4N\left[\sum_{s=0}^{1}\,\frac{1}{s!}\left(\begin{array}{c} 1 \\
s\end{array}\right)\,\left(-\frac{\lambda_{0,k}}{4N}\right)^{s}\,-\,1\right]
= 4N\,\left(-\frac{\lambda_{0,k}}{4N}\right)\nonumber \\ &=&
-\lambda_{0,k}=-\lambda_{0,k}\,
\langle\psi_{0}|\hat{\Phi}_{0,k}^{\left(N\right)}|\psi_{0}\rangle
\label{diagoeigf0} \ . \eeqa Hence the fuzzy Laplacian maps the
element $\hat{\Phi}_{0,k}^{\left(N\right)}$ into another diagonal
element whose coefficient, respect to
$|\psi_{0}\rangle\langle\psi_{0}|$, is just  a multiplication by
$-\lambda_{0,k}$ of the coefficient of
$\hat{\Phi}_{0,k}^{\left(N\right)}$ respect to the same basis
element. Proceeding along this way we project the element
$\nabla^{2}_{\left(N\right)}\left(\hat{\Phi}_{n,k}^{\left(N\right)}\right)$
on $|\psi_{b}\rangle\langle\psi_{b}|$, with $b=1,\ldots,N-1$:
\beq \langle\psi_{b}|\
\nabla^{2}_{\left(N\right)}
\left(\hat{\Phi}_{0,k}^{\left(N\right)}\right)|\psi_{b}\rangle=
\left(b+1\right)\left(\phi_{0,k}^{\left(b+1,N\right)}\,-\,
\phi_{0,k}^{\left(b,N\right)}\right)\,+
\,b\left(\phi_{0,k}^{\left(b-1,N\right)}\,-\,\phi_{0,k}^{\left(b,N\right)}\right)
\label{lapladiagoprob} \ . \eeq To study the r.h.s.\ of this
relation, we need to calculate the difference: \beqa
\Phi_{0,k}^{\left(b+1,N\right)}-\Phi_{0,k}^{\left(b,N\right)}&=&
\left[\sum_{s=0}^{b+1}\,\left(-\frac{\lambda_{0,k}}{4N}\right)^{s}
\frac{1}{s!}\left(\begin{array}{c} b+1 \\ s
\end{array}\right)\right]
\,-\,\left[\sum_{s=0}^{b}\,\left(-\frac{\lambda_{0,k}}{4N}\right)^{s}
\frac{1}{s!}\left(\begin{array}{c} b \\ s
\end{array}\right)\right]
\nonumber \\ &=&
\sum_{s=1}^{b+1}\,\left[\frac{1}{s!}\,\left(\begin{array}{c} b \\
s-1
\end{array}\right)\right]\,\left(-\frac{\lambda_{0,k}}{4N}\right)^{s}
\label{phibpiuunomenob} \ ,
\eeqa
hence the equation~(\ref{lapladiagoprob}) becomes:
\beqa
\langle\psi_{b}|\
\nabla^{2}_{\left(N\right)}\left(\hat{\Phi}_{0,k}^{\left(N\right)}
\right)|\psi_{b}\rangle&=&
4N\left(\left(b+1\right)\left(\phi_{0,k}^{\left(b+1,N\right)}-
\phi_{0,k}^{\left(b,N\right)}\right)\,+
\,b\left(\phi_{0,k}^{\left(b-1,N\right)}-\phi_{0,k}^{\left(b,N\right)}
\right)\right) \nonumber
\\ &=&
4N\left(\sum_{s=1}^{b+1}\,\left[\frac{b+1}{s!}\,\left(\begin{array}{c}
b
\\ s-1 \end{array}
\right)\right]\,\left(-\frac{\lambda_{0,k}}{4N}\right)^{s} \right .+\nonumber\\
&&\left.-\sum_{s=1}^{b}\,\left[\frac{b}{s!}\,\left(\begin{array}{c}
b-1
\\ s-1
\end{array}\right)\right]\left(-\frac{\lambda_{0,k}}{4N}\right)^{s}\right)
\nonumber\\ &=& -\lambda_{0,k}\,
\left[\,\sum_{s=0}^{b}\,\frac{1}{s!}\,\left(\begin{array}{c} b
\\ s \end{array}\right)\,
\left(-\frac{\lambda_{0,k}}{4N}\right)^{s}\,\right] =
-\lambda_{0,k}\,\Phi_{0,k}^{\left(b,N\right)}\label{diagoeigfb} \
,
\eeqa
where to reach the expression on the last line we have recollected
equal powers of the $\lambda$ and made some simplifications. This
relation is similar to~(\ref{diagoeigf0}). Whatever
$\lambda_{0,k}$, the fuzzy Laplacian acts on the first $N$
components of the fuzzy elements
$\hat{\Phi}_{0,k}^{\left(N\right)}$ simply as a multiplication by
$-\lambda_{0,k}$. To conclude that the elements
$\hat{\Phi}_{0,k}^{\left(N\right)}$ are eigenstates of the fuzzy
Laplacian we need to study the last component projection of
$\nabla^{2}_{\left(N\right)}\left(\hat{\Phi}_{0,k}^{\left(N\right)}\right)$,
namely that on the basis element
$|\psi_{N}\rangle\langle\psi_{N}|$. Moreover, we have not yet
obtained any conditions on the possible eigenvalues of the fuzzy
Laplacian.

Let us project:
\beq
\langle\psi_{N}|\
\nabla^{2}_{\left(N\right)}\left(\hat{\Phi}_{0,k}^{\left
(N\right)}\right)|\psi_{N}\rangle
=4N\left(N\Phi_{0,k}^{\left(N-1,N\right)}-
\left(2N+1\right)\Phi_{0,k}^{\left(N,N\right)}\right) \ ,
\eeq
and impose:
\beq
\langle\psi_{N}|\
\nabla^{2}_{\left(N\right)}\left(\hat{\Phi}_{0,k}^{\left(
N\right)}\right)|\psi_{N}\rangle
=-\lambda_{0,k}\Phi_{0,k}^{\left(N,N\right)}=
-\lambda_{0,k}\langle\psi_{N}|\hat{\Phi}_{0,k}^{\left(N\right)}
|\psi_{N}\rangle \label{diagoeigfN} \ . \eeq This can be cast in
the form:
\bea
\sum_{s=0}^{N}\,\frac{1}{s!}\left(\begin{array}{c} N
\\ s
\end{array}\right)\left(-\frac{\lambda_{0,k}}{4N}\right)^{s+1}+
\left(N+1\right)\sum_{s=0}^{N}\frac{1}{s!}\,\left(\begin{array}{c}
N
\\ s
\end{array}\right)\left(-\frac{\lambda_{0,k}}{4N}\right)^{s}&&\nonumber\\+
N\,\sum_{s=0}^{N}\frac{1}{s!}\,\left(\begin{array}{c} N-1 \\ s-1
\end{array}\right)\left(-\frac{\lambda_{0,k}}{4N}\right)^{s}&=&0.
\eea
Recollecting terms with equal powers of $\lambda$, the equation
can be written as:
\beq E_{N}\left(\frac{\lambda_{0,k}}{4N}\right)\,
\equiv\,\sum_{a=0}^{N+1}\,\frac{\left(-1\right)^{a+N+1}}{\left(N+1-a\right)!}
\,\left[\frac{\left(N+1\right)!}{a!}\right]^{2}\,
\left(\frac{\lambda_{0,k}}{4N}\right)^{a} =0 \label{eqEN}\ . \eeq
The meaning of this calculation is clear:
relation~(\ref{diagoeigfN}) is not identically satisfied by all
value of $\lambda_{0,k}$. So it becomes an equation, giving the
admissible eigenvalues $\lambda_{0,k}^{\left(N\right)}$ of the
fuzzy Laplacian on the subspace of diagonal fuzzy elements in
$\mathcal{A}^{\left(N\right)}_{\theta}$: \beq
\nabla^{2}_{\left(N\right)}\left(\hat{\Phi}_{0,k}^{\left(N\right)}\right)\,=
\,-\lambda_{0,k}^{\left(N\right)}\hat{\Phi}^{\left(N\right)}_{0,k}
\label{eigeqdiagonal}\eeq if and only if: \beq
E_{N}\left(\frac{\lambda_{0,k}^{\left(N\right)}}{4N}\right)=0\label{eigeqdiagoEN}
\ . \eeq

We now show that the solutions of this last equation are actually
the eigenvalues of the fuzzy Laplacian (in the $n=0$ case). From
the action of $\nabla^2_{(N)}$ on a generic element of this
diagonal subspace, for
$\hat{\Phi}^{\left(N\right)}=\sum_{a=0}^{N}\,
\phi_{a}^{\left(N\right)}\,|\psi_{a}\rangle\langle\psi_{a}|$ we
have:
\beqa
\nabla^{2}_{\left(N\right)}\hat{\Phi}^{\left(N\right)}&=&
4N\Big(\left(\phi_{1}^{\left(N\right)}-\phi_{0}^{\left(N\right)}\right)
|\psi_{0}\rangle\langle\psi_{0}| +
\left(N\phi_{N-1}^{\left(N\right)}-
\left(2N+1\right)\phi_{N}^{\left(N\right)}\right)|\psi_{N}\rangle
\langle\psi_{N}|+ \nonumber\\
&&+\sum_{a=1}^{N-1}\left[\left(a+1\right)\left(\phi_{a+1}^{\left(N\right)}
-\phi_{a}^{\left(N\right)}\right)+
a\left(\phi_{a-1}^{\left(N\right)}-
\phi_{a}^{\left(N\right)}\right)\right]
\,|\psi_{a}\rangle\langle\psi_{a}|\Big) \label{diagolapl} \ ,
\eeqa the eigenvalue equation
(\ref{eigeqdiagonal}) can be written in components as:  \beqa
\phi_{1}^{\left(N\right)}-\phi_{0}^{\left(N\right)}&=&
-\left(\frac{\lambda_{0,k}^{\left(N\right)}}{4N}\right)\,
\phi_{0}^{\left(N\right)} \nonumber \\
\left(a+1\right)\left(\phi_{a+1}^{\left(N\right)}-
\phi_{a}^{\left(N\right)}\right)+a
\left(\phi_{a-1}^{\left(N\right)}-\phi_{a}^{\left(N\right)}\right)&=&
-\left(\frac{\lambda_{0,k}^{\left(N\right)}}{4N}\right)\,
\phi_{a}^{\left(N\right)}
\,\,\,\,\,\,\,\,\,\,\,\,\,\mbox{if}\,a\neq 0,N\nonumber \\
N\phi_{N-1}^{\left(N\right)}-\left(2N+1\right)\phi_{N}^{\left(N\right)}&=&
-\left(\frac{\lambda_{0,k}^{\left(N\right)}}{4N}\right)\,
\phi_{N}^{\left(N\right)} \label{eigeqdiagocomp}\ . \eeqa This set
of equations can be given a matrix form. If we set
$x_N=\lambda_{0,k}^{\left(N\right)}/4N$, then the eigenvalues are
related to the roots of the characteristic polynomial
$P_{N}\left(x_N\right)$, given by the determinant of the matrix
\beq
P_{N}=\det\,\left(
 {\textstyle
\begin{array}{ccccc}
x_N-1 & 1 & \ldots & 0 & 0 \\
 1 & x_N-3 & \ldots & 0 & 0 \\
 0 & 2 & \ldots & 0 & 0 \\
\vdots & \vdots & \ddots & \vdots &
 \vdots \\
 0 & 0 & \ldots & x_N-\left(2N-1\right)
& N \\ 0 & 0 & \ldots & N & x_N-\left(2N+1\right)
\end{array}
} \right)\label{secdiagomat}\ . \eeq The eigenvalue equation for
the fuzzy Laplacian in the diagonal matrices subspace is:
\beq
P_{N}\left(\frac{\lambda_{0,k}^{\left(N\right)}}{4N}\right)=0
\label{eigeqpolydiago} \ . \eeq Matrices of the kind
of~\eqn{secdiagomat} ara called Jacobi matrix~\cite{Gantmacher}, and
their characteristic polynomial is given inductively:
\beq
P_{N}\left(x_N\right)=\left(x_N-\left(2N+1\right)\right)\,
P_{N-1}\left(x_N\right)\,-\,N^{2}\,P_{N-2}\left(x_N\right)
\label{detdiago} \ . \eeq Equation~(\ref{eigeqpolydiago}) is
equivalent to~(\ref{eigeqdiagoEN}). This result, proven in the
appendix, closes the analysis on the spectral properties for the
$n=0$ case: eigenstates are given by
$\hat{\Phi}_{0,k}^{\left(N\right)}$, while eigenvalues are given
by the roots of~(\ref{eigeqpolydiago}).

We approach the problem for the action of the fuzzy Laplacian in
the nondiagonal subspaces of
$\mathcal{A}^{\left(N\right)}_{\theta}$, defined by the integer
$n$, following a similar strategy. The dimension of a generic
subspace is $N+1-n$, the basis elements are
$|\psi_{a}\rangle\langle\psi_{a+n}|$ with $a=0\,\ldots,\,N-n$ so
the fuzzy Laplacian can be seen as an operator acting on
$\complex^{N+1-n}$. Going through the path described for the $n=0$
case, we prove that the eigenvalue problem in this subspace is
solved by the fuzzy elements $\hat{\Phi}^{\left(N\right)}_{n,k}$
(\ref{Phinfuzz}) with $k=1,\,\ldots\,N+1-n$: \beq
\nabla^{2}_{\left(N\right)}\,\hat{\Phi}_{n,k}^{\left(N\right)}
=-\lambda_{n,k}^{\left(N\right)}\,
\hat{\Phi}_{n,k}^{\left(N\right)}\label{eigeqn}\eeq if and only if
the eigenvalues are solutions of: \beq E_{N}^{\left(n\right)}=
\sum_{k=0}^{N-n+1}\,\left(-1\right)^{k+N-n+1}\,\frac{\left
(N+1\right)!\left(N-n+1\right)!}
{k!\left(n+k\right)!\left(N-n-k+1\right)!}\,
\left(\frac{\lambda_{n,k}^{\left(N\right)}}{4N}\right)^{k}=0
\label{eqENn}\ . \eeq This equation generalises the equation
(\ref{eqEN}), reducing to that for $n=0$.

The proof will closely follow the path outlined in the diagonal
case. First project the image element
$\nabla^{2}_{\left(N\right)}\left(\hat{\Phi}_{n,k}^{\left(N\right)}\right)$
on the first basis element $|\psi_{0}\rangle\langle\psi_{n}|$.
Following the general case~(\ref{basislapl}), one can see that:
\beq
\la\psi_{0}|\left[\nabla^{2}_{\left(N\right)}
\left(|\psi_{a}\rangle\langle\psi_{a+n}|\right)\right]|\psi_{n}\rangle
=4N\left[\delta_{1,a}\sqrt{n+1}-\left(n+1\right)\delta_{0,a}\right]
\ . \eeq After some algebra using the definition~(\ref{Phinfuzz}):
\be
\langle\psi_{0}|\left[\nabla^{2}_{\left(N\right)}\hat{\Phi}_{n,k}^{\left
(N\right)}\right]|\psi_{n}\rangle
=-\lambda_{n,k}
\langle\psi_{0}|\hat{\Phi}_{n,k}^{\left(N\right)}|\psi_{n}\rangle
\ee
where the value $\lambda_{n,k}$ is again just a parameter.

Now we can project the
$\nabla^{2}_{\left(N\right)}\left(\hat{\Phi}_{n,k}^{\left(N\right)}\right)$
element on the basis elements
$|\psi_{b}\rangle\langle\psi_{b+n}|$, with $b\neq 0,N-n$. Again, 
after some straightforward algebra:
\be
\langle\psi_{b}|\left[\nabla^{2}_{\left(N\right)}\hat{\Phi}_{n,k}^{\left(
N\right)}\right]|\psi_{b+n}\rangle
= -\lambda_{n,k}\,
\langle\psi_{b}|\hat{\Phi}_{n,k}^{\left(N\right)}|\psi_{b+n}\rangle.
\ee
Projecting the
$\nabla^{2}_{\left(N\right)}\left(\hat{\Phi}_{n,k}^{\left(N\right)}\right)$
element on the last basis element:
\beqa
&\langle\psi_{N-n}|\left[\nabla^{2}_{\left(N\right)}
\hat{\Phi}_{n,k}^{\left(N\right)}\right]|\psi_{N}\rangle=&\nonumber\\&
=4N\,\left(\frac{\lambda_{n,k}}{4N}\right)^{n/2}\left(
\sum_{s=0}^{N-n-1}\,\frac{\sqrt{\left(N-n\right)!N!}}
{s!\left(n+s\right)!\left(N-n-1-s\right)!}
\left(-\frac{\lambda_{n,k}}{4N}\right)^{s}\,+\right.&\nonumber
\\  &- \left.
\sum_{s=0}^{N-n}\,\left(2N-n+1\right)
\frac{\sqrt{\left(N-n\right)!N!}}{s!\left(n+s\right)!\left(N-n-s\right)!}
\left(-\frac{\lambda_{n,k}}{4N}\right)^{s}\right)&\ . \eeqa This
r.h.s.\ defined to be:
\be
 -\lambda_{n,k}\,
\langle\psi_{N-n}|\hat{\Phi}_{n,k}^{\left(N\right)}|\psi_{N}\rangle
=-\lambda_{n,k}\,
\left(\frac{\lambda_{n,k}}{4N}\right)^{n/2}\left(
\sum_{s=0}^{N-n}\,\frac{\sqrt{\left(N-n\right)!N!}}
{s!\left(n+s\right)!\left(N-n-s\right)!}
\left(-\frac{\lambda_{n,k}}{4N}\right)^{s}\right)
\ee
This is an equation for  $\lambda_{n,k}$, that can
be cast exactly in the form of~(\ref{eqENn}):
\beq
E_{N}^{\left(n\right)}=
\sum_{k=0}^{N-n+1}\,\left(-1\right)^{k+N-n+1}\,
\frac{\left(N+1\right)!\left(N-n+1\right)!}
{k!\left(n+k\right)!\left(N-n-k+1\right)!}\,
\left(\frac{\lambda_{n,k}}{4N}\right)^{k}=0\ . \eeq So the
eigenvalues $\lambda_{n,k}^{\left(N\right)}$ of the fuzzy
Laplacian in this stability subspace are given by the solution of:
\beq
E_{N}^{\left(n\right)}\left(
\frac{\lambda_{n,k}^{\left(N\right)}}{4N}\right)=0\label{eigeqENn}
\ . \eeq

It is again possible to prove that this equation is exactly
equivalent to the secular equation coming from the matrix
representation of the action of the fuzzy Laplacian on each of
this $N-n+1$ dimensional subspaces. The analogue of the matrix
(\ref{secdiagomat}) is, in this case:
\beq
P_{N}^{\left(n\right)}=\det\left({\small
\begin{array}{ccccc}
 x_N-\left(n+1\right) & \sqrt{\left(n+1\right)}& \ldots & 0 & 0
\\
 \sqrt{\left(n+1\right)} & x_N-\left(n+3\right) &  \ldots & 0 & 0
\\
\vdots &  \vdots &  \ddots & \vdots & \vdots
\\
0 & 0 &  \ldots & x_N-\left(2N-n-1\right) &
\sqrt{N\left(N-n\right)}
\\
0 & 0 &   \ldots & \sqrt{N\left(N-n\right)} &
x_N-\left(2N-n+1\right)
\end{array}
} \right) \ . \label{secnmat}  \eeq The characteristic polynomial
is of degree is $N+1-n$ and a recursive relation holds, analogous
to~(\ref{detdiago}):
\beq
P_{N}^{\left(n\right)}\left(x_N\right)=
\left(x_N-\left(2N-n+1\right)\right)
P_{N-1}^{\left(n\right)}\left(x_N\right)
-N\left(N-n\right)\,P_{N-2}^{\left(n\right)}\left(x_N\right)
\label{detn} \ . \eeq In the appendix there is the proof of the
equivalence of the relation
\beq
P_{N}^{\left(n\right)}\left(x_N\right)=0 \label{eigeqpolyn}
\eeq
with~(\ref{eigeqENn}). Setting $n=0$ the diagonal case is
recovered.

The analysis for the case $n<0$ is straightforward. The eigenvalue
equation is the same, and this shows why there is a set of doubly
degenerate eigenvalues. The eigenstates are obtained by those for
positive $n$ by complex conjugation. This completes the spectral
analysis of the fuzzy laplacian. Since we have proved that the
eigenstates of this fuzzy Laplacian are the fuzzified version of
the continuum eigenfunctions, we call
$\hat{\Phi}_{n,k}^{\left(N\right)}$ \emph{fuzzy Bessels}.

\subsection{Comparison of Bessel Functions}
\begin{figure}[htbp]
\epsfxsize=2.3 in \centerline{\epsfxsize=2.2
in\epsffile{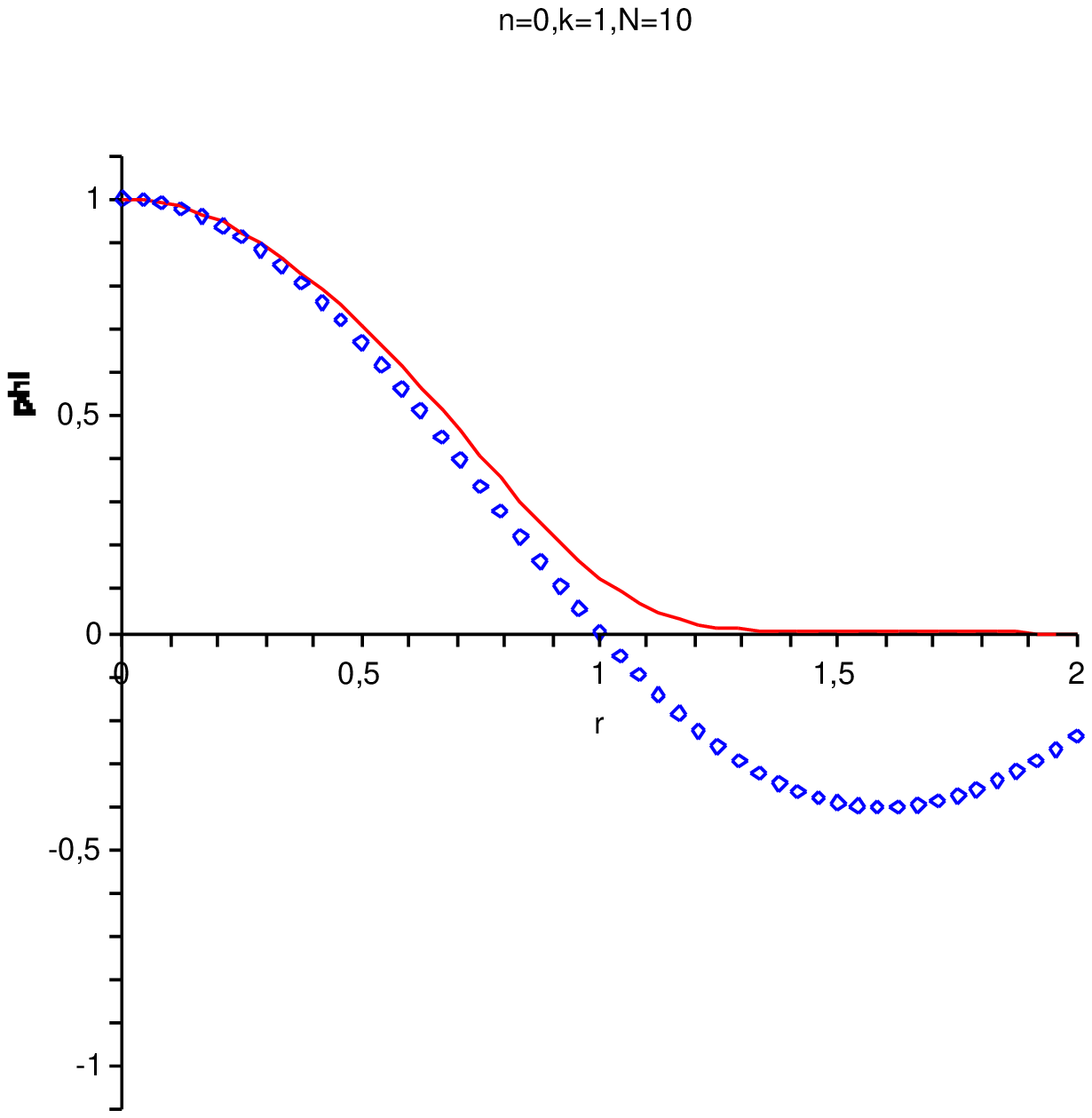}\epsfxsize=2.2
in\epsffile{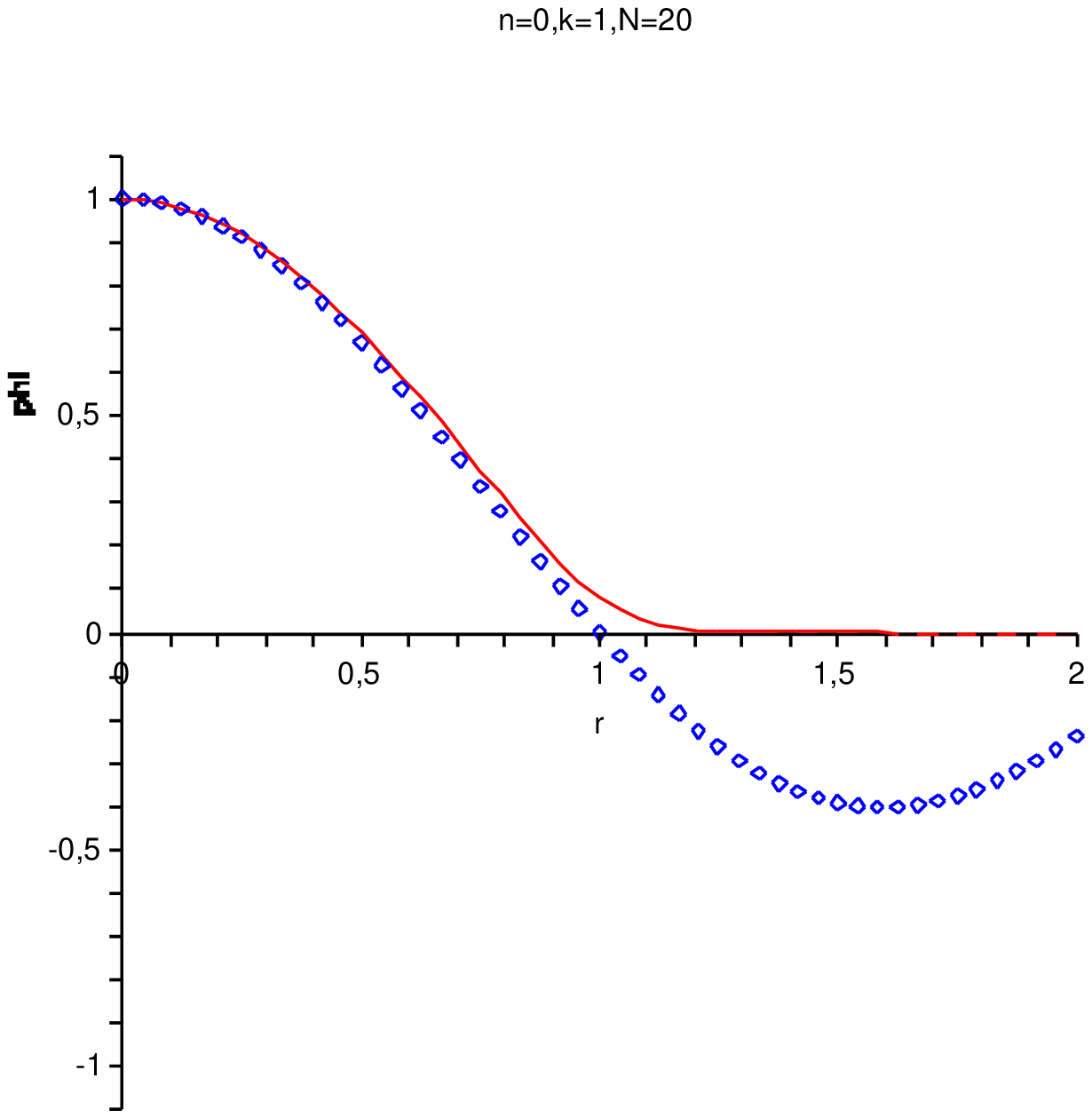}\epsfxsize=2.2
in\epsffile{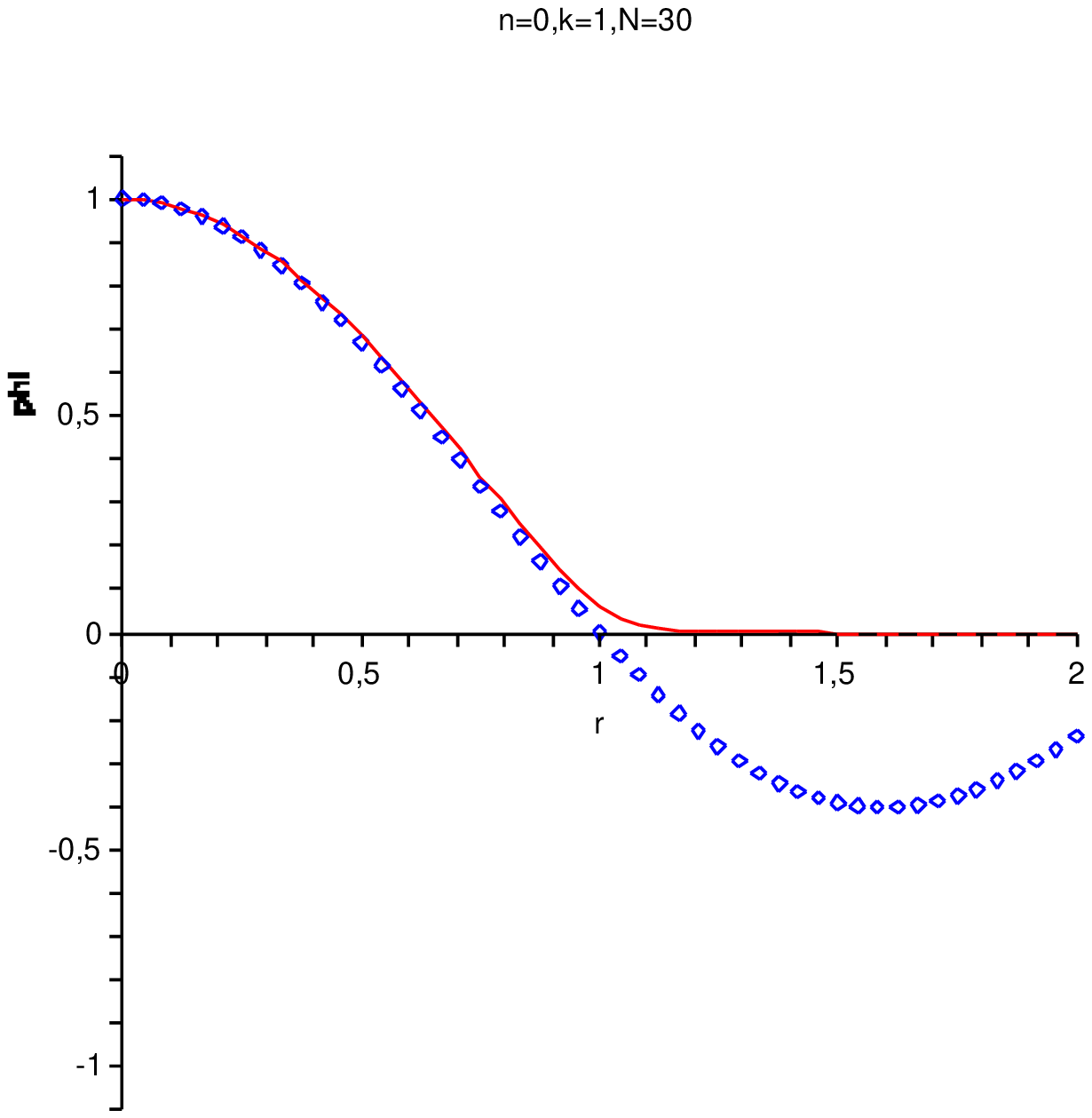}} \caption{\baselineskip=12pt {\it
Comparison of the radial shape for the symbol
$\Phi^{\left(N\right)}_{0,1}\left(r,\varphi\right)$ (continuum
line), the symbol of the eigenmatrix of the fuzzy Laplacian for
$N=10,20,30$, with $\Phi_{0,1}\left(r,\varphi\right)$.}}
\label{zero_uno}
\end{figure}

\begin{figure}[htbp] \epsfxsize=2.3 in \centerline{\epsfxsize=2.2
in\epsffile{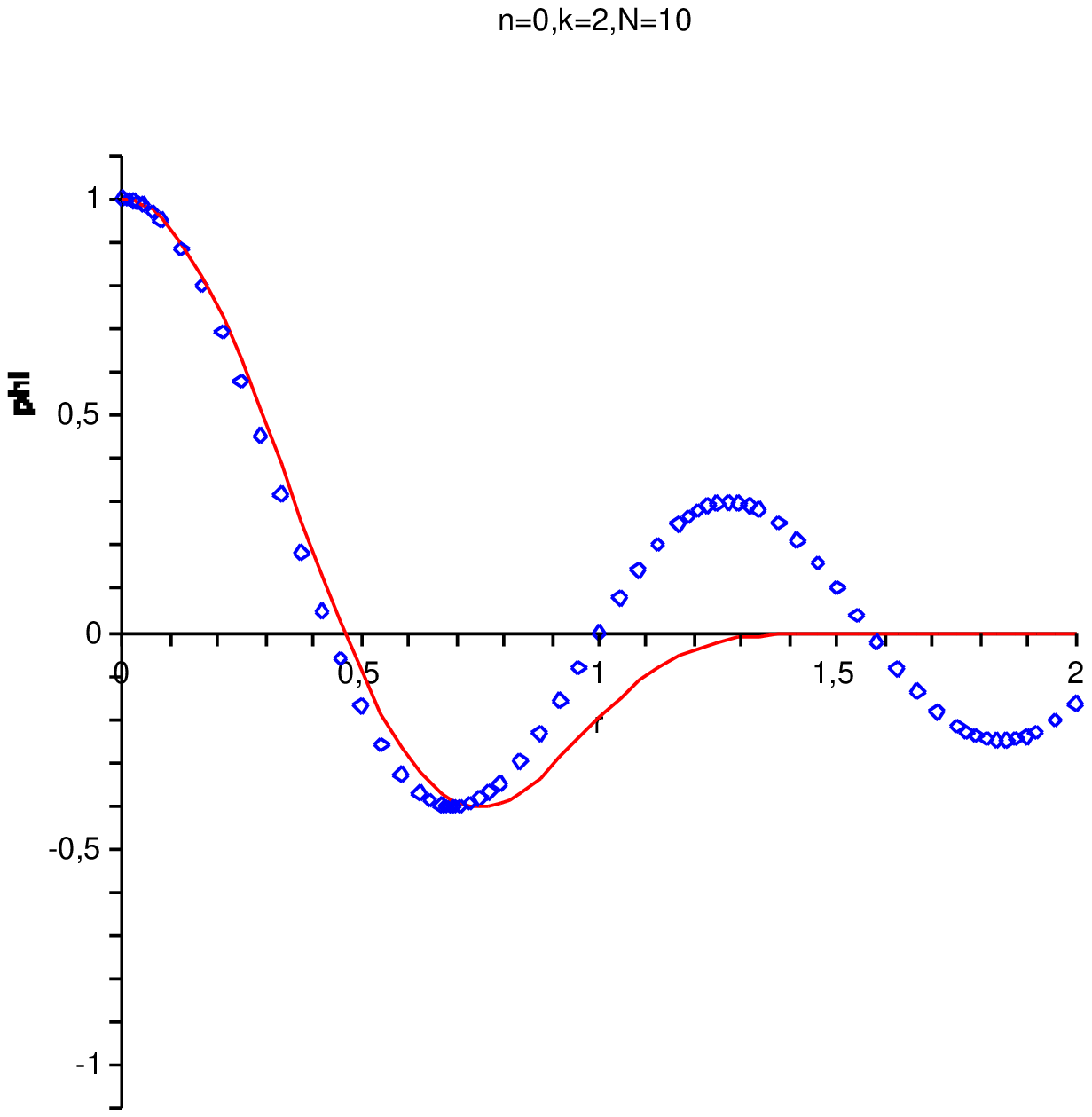}\epsfxsize=2.2
in\epsffile{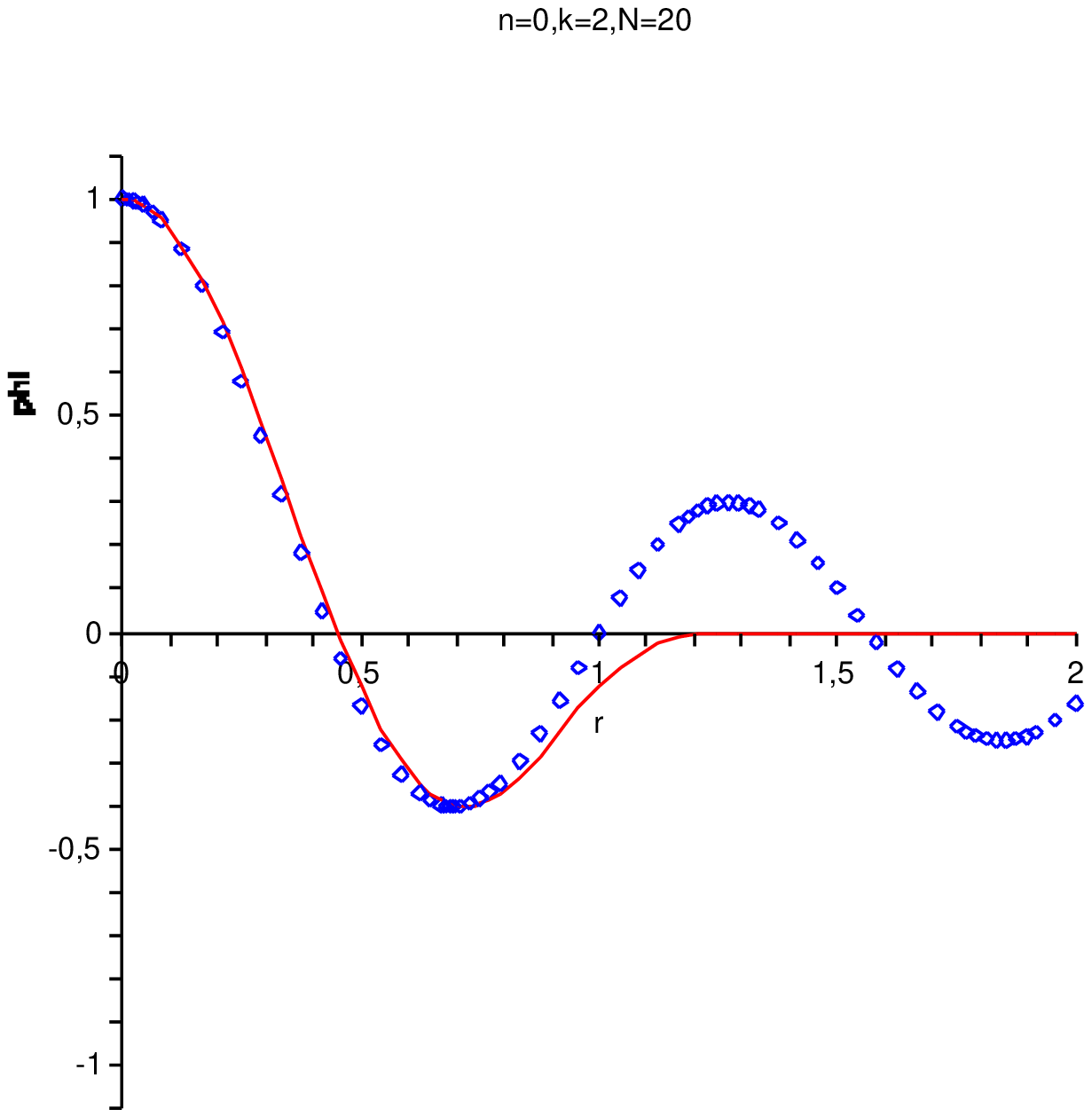}\epsfxsize=2.2
in\epsffile{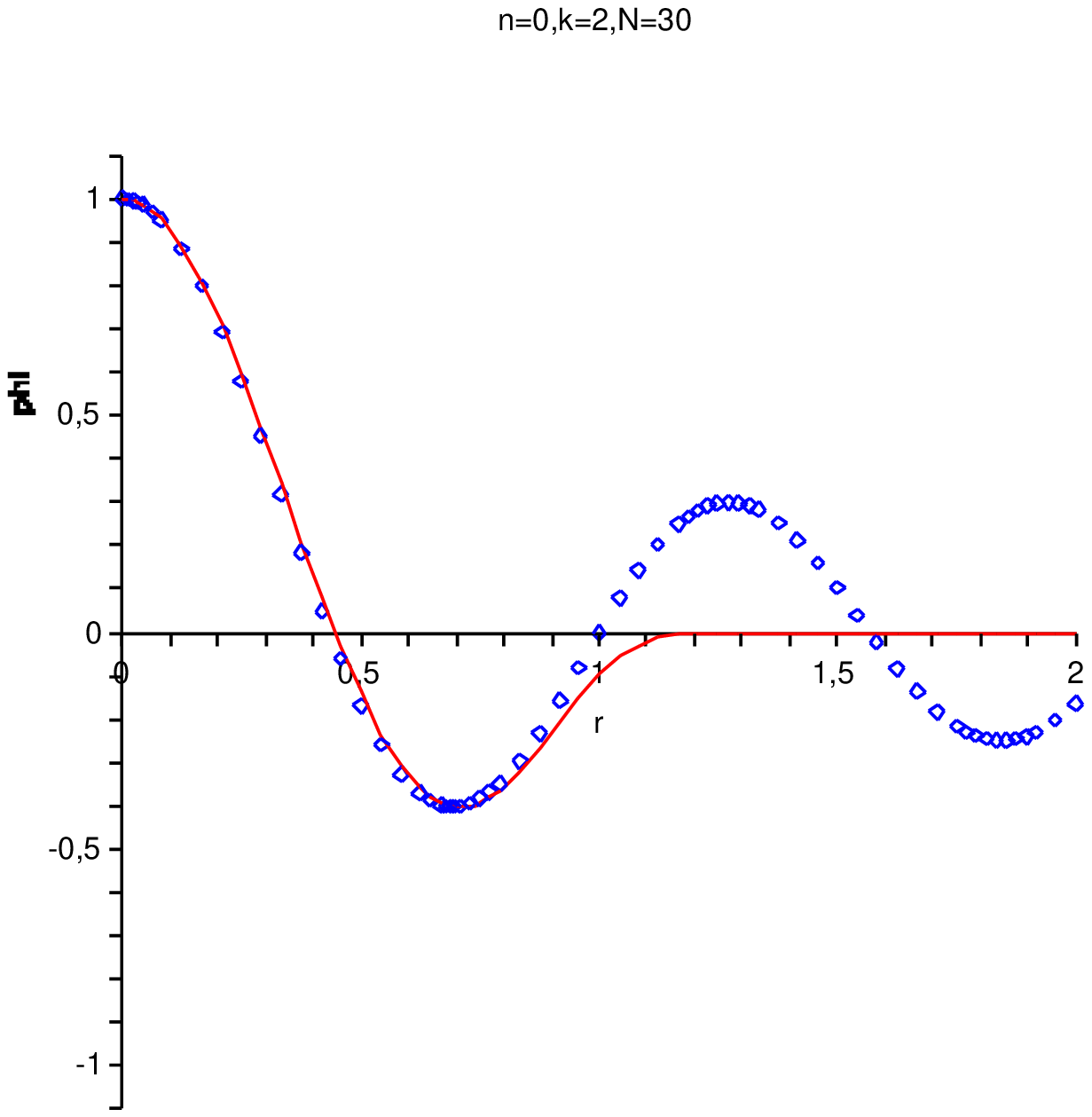}} \caption{\baselineskip=12pt {\it
Comparison of the radial shape for the symbol
$\Phi_{0,2}^{\left(N\right)}\left(r\right)$ (continuum line) with
$\Phi_{0,2}\left(r\right)$ for $N=10,20,30$.} } \label{zero_due}
\end{figure}

\begin{figure}[htbp]
\epsfxsize=2.3 in \centerline{\epsfxsize=2.2
in\epsffile{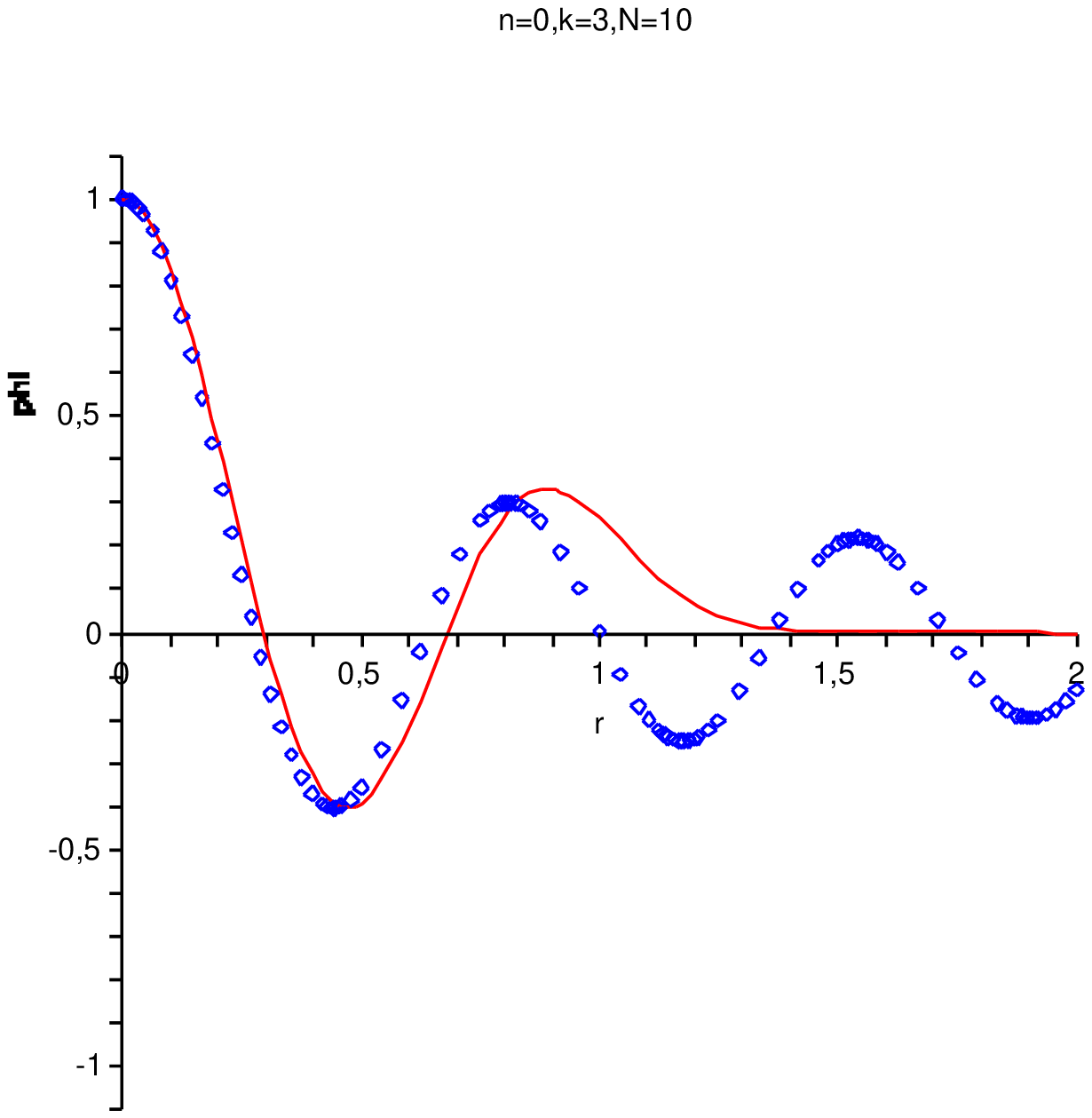}\epsfxsize=2.2
in\epsffile{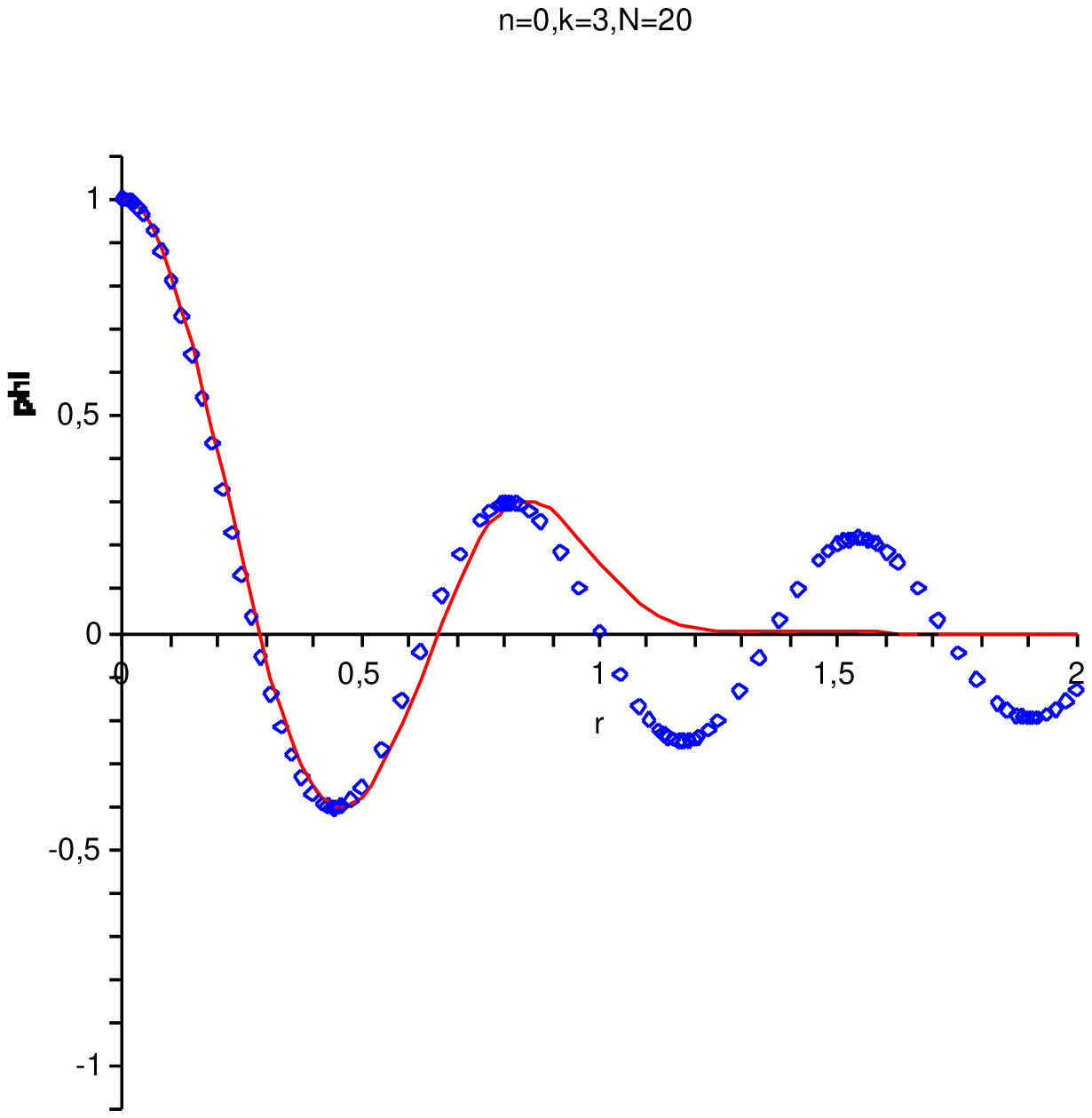}\epsfxsize=2.2
in\epsffile{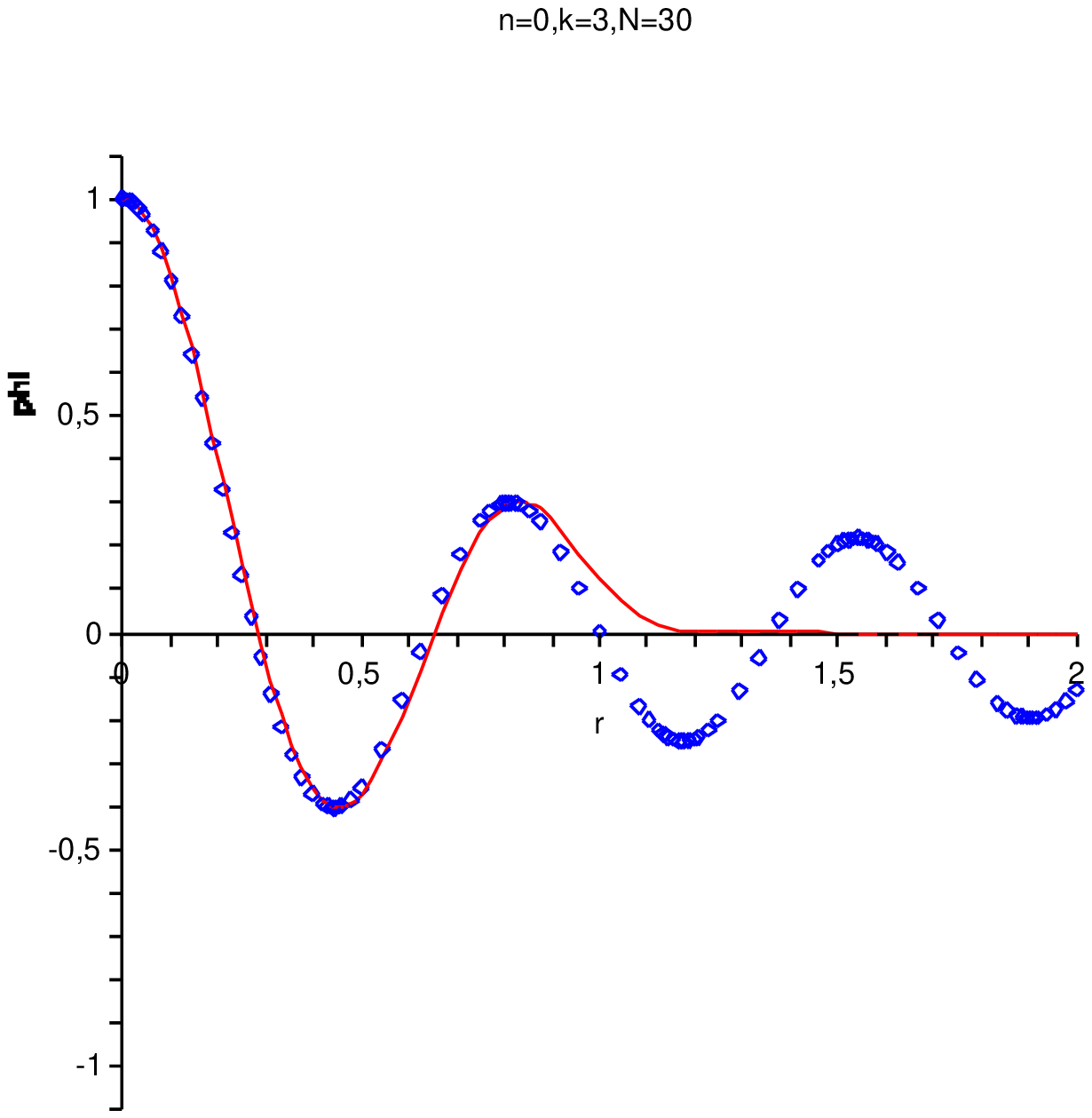}} \caption{\baselineskip=12pt {\it
Comparison of the radial shape for the symbol
$\Phi_{0,3}^{\left(N\right)}\left(r\right)$ (continuum line) with
$\Phi_{0,3}\left(r\right)$ for $N=10,20,30$.}}
\bigskip
\label{zero_tre}
\end{figure}

\begin{figure}[htbp]
\epsfxsize=2.3 in \centerline{\epsfxsize=2.2
in\epsffile{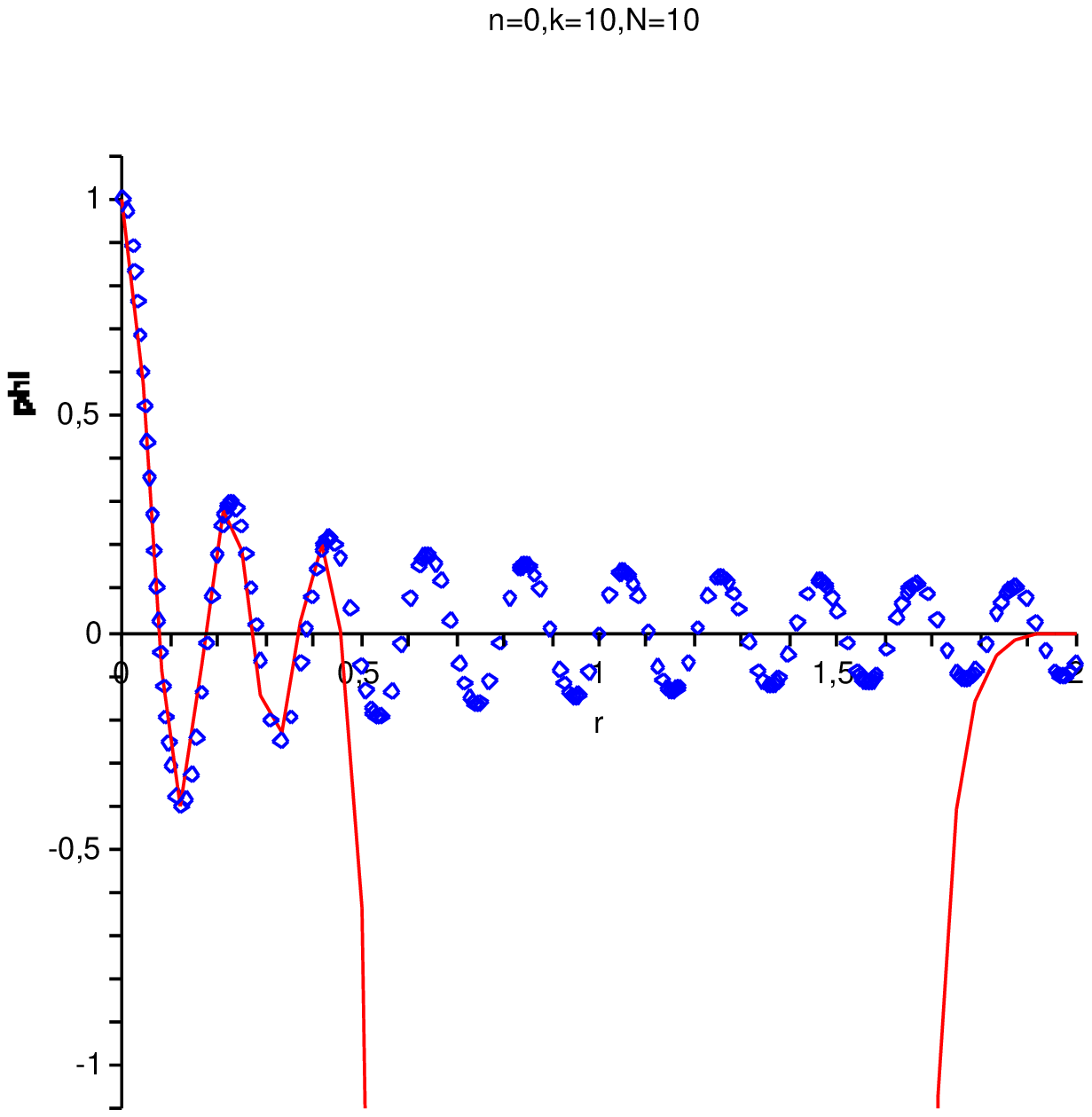}\epsfxsize=2.2
in\epsffile{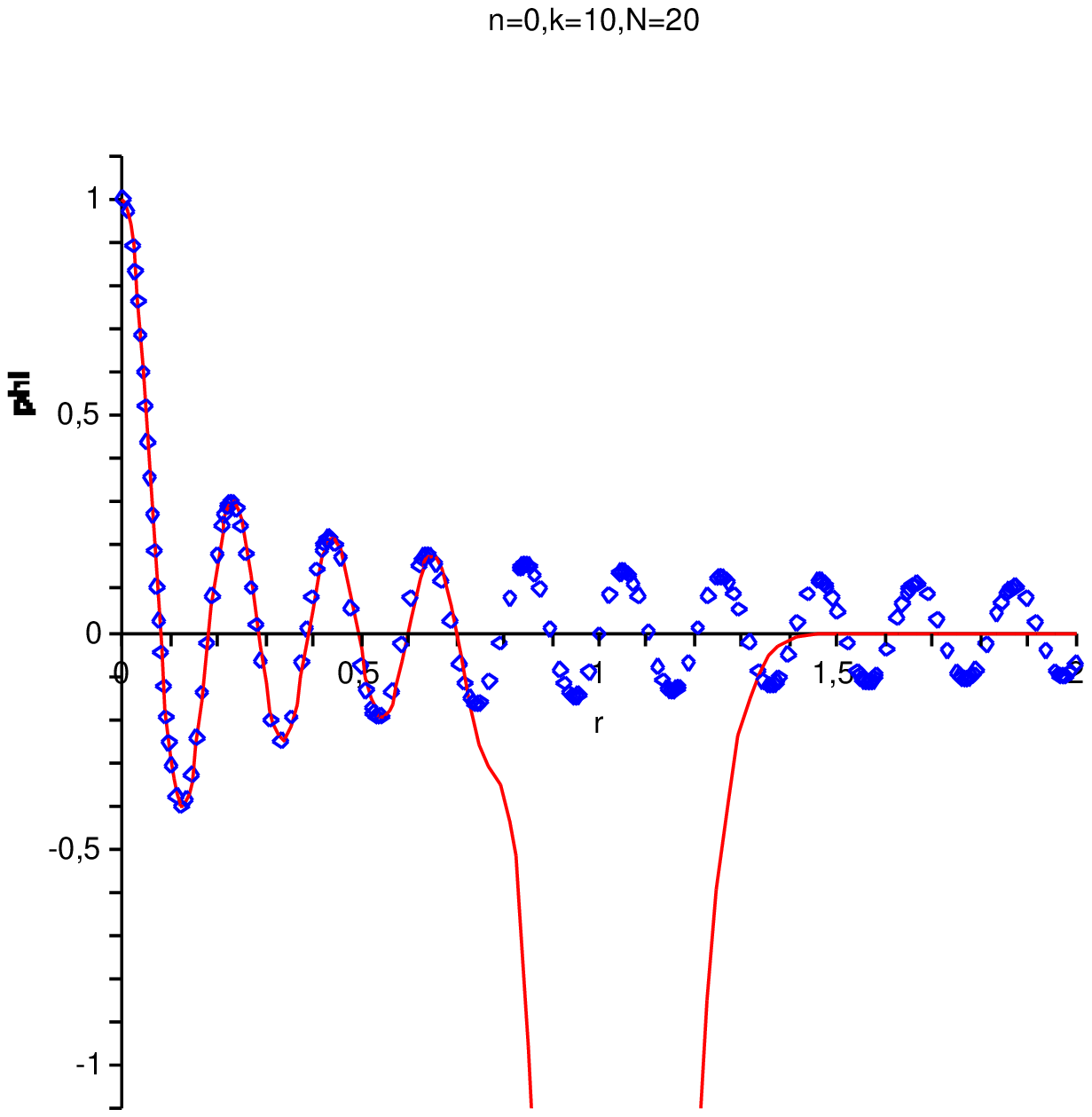}\epsfxsize=2.2
in\epsffile{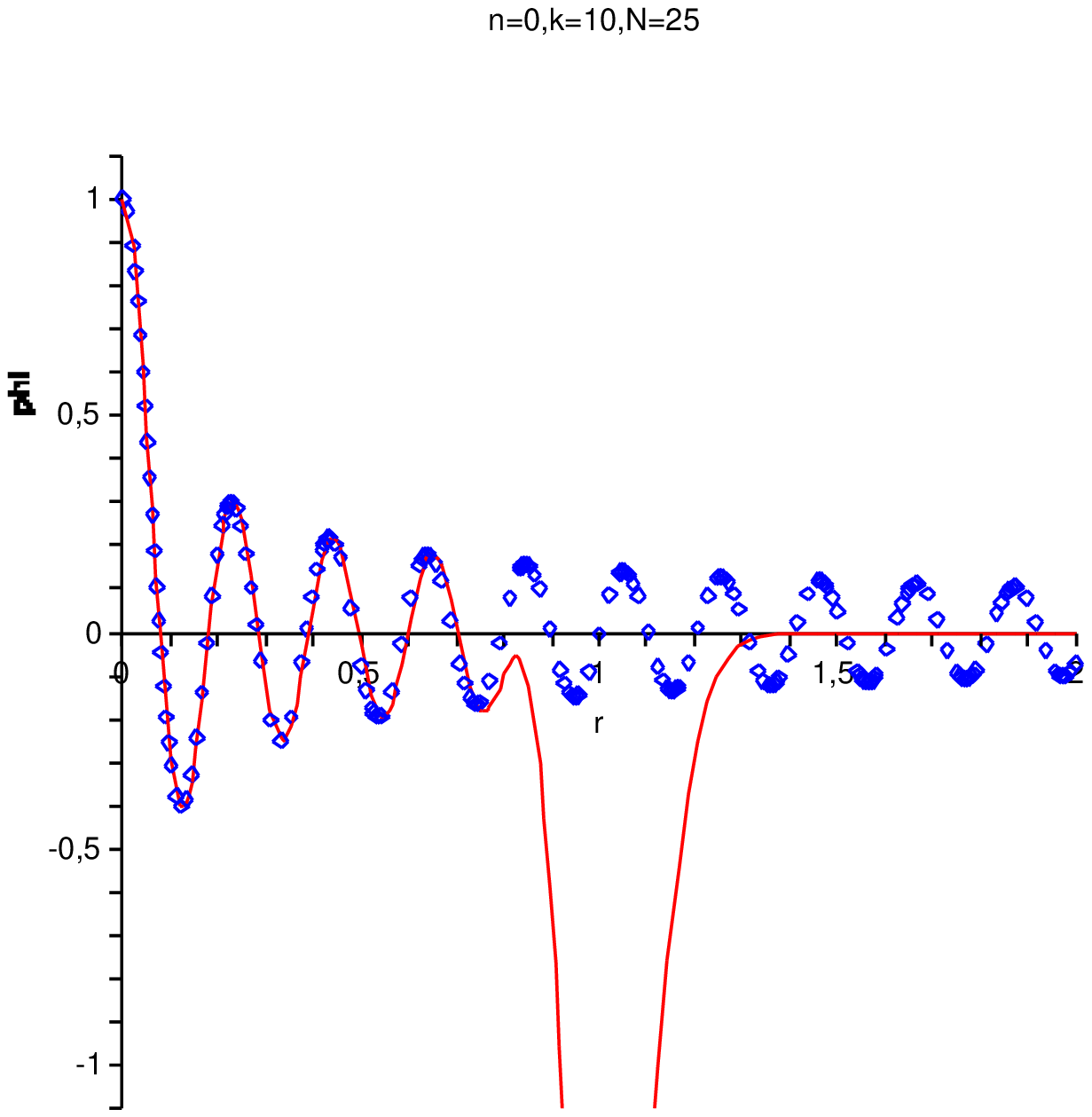}} \caption{\baselineskip=12pt {\it
Comparison of the radial shape for the
$\Phi_{0,10}^{\left(N\right)}$ symbol (continuum line)for
$N=10,20,25$. The bump goes out of the picture. As $N$ increases, 
it becomes narrower and narrower. }}
\bigskip
\label{zero_dieci_prima}
\end{figure}

\begin{figure}[htbp]
\epsfxsize=2.3 in \centerline{\epsfxsize=2.2
in\epsffile{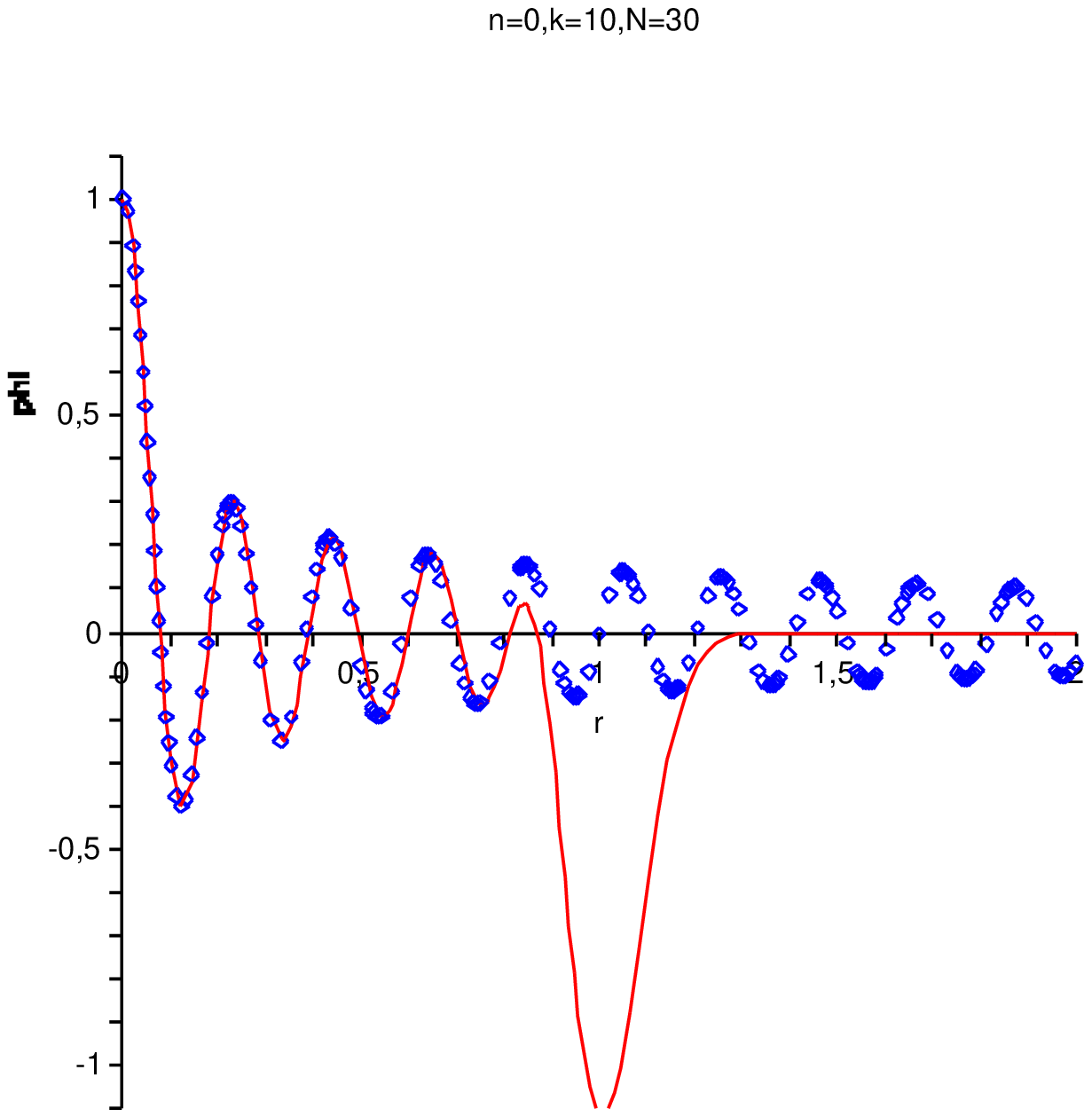}\epsfxsize=2.2
in\epsffile{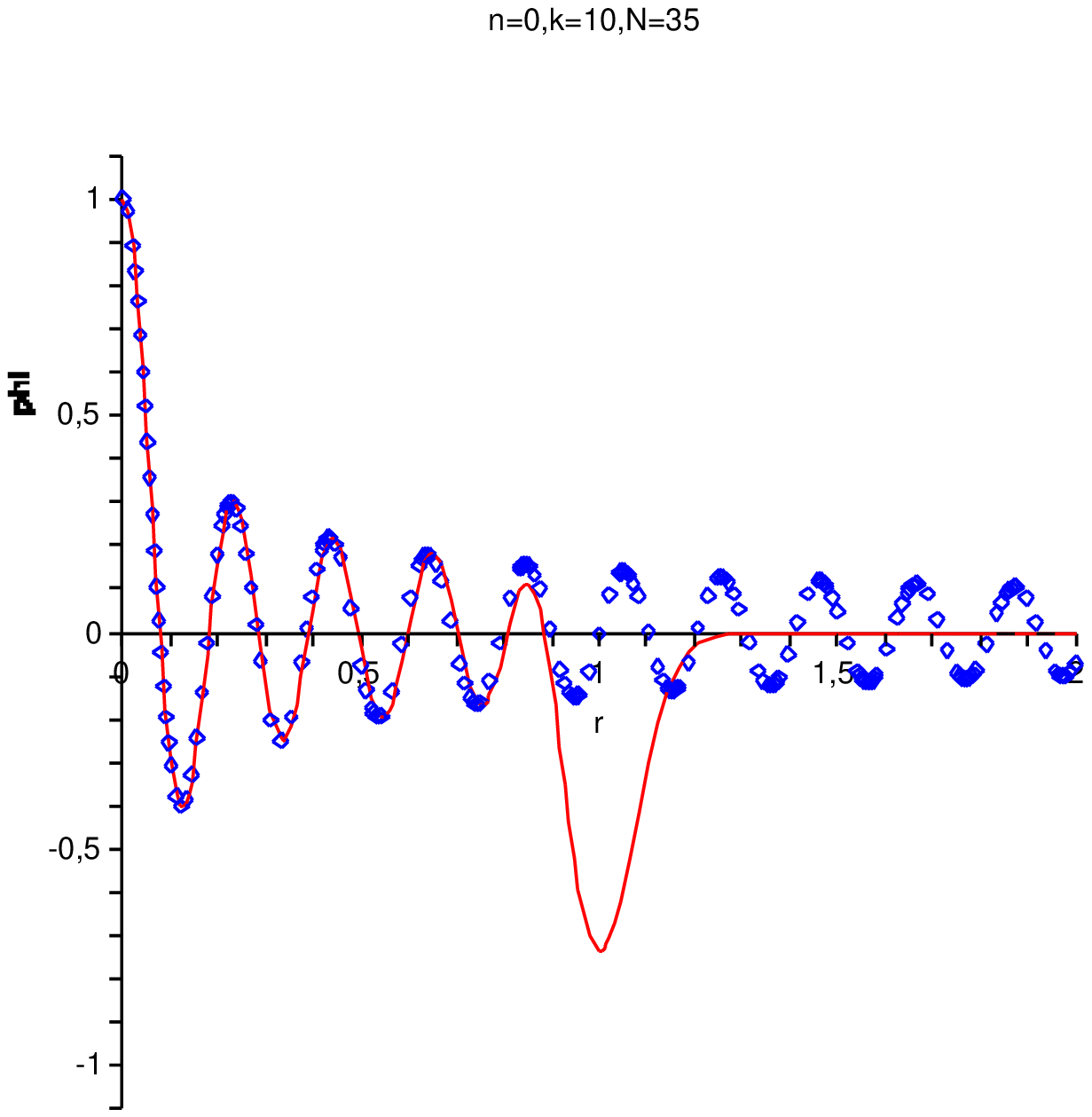}\epsfxsize=2.2
in\epsffile{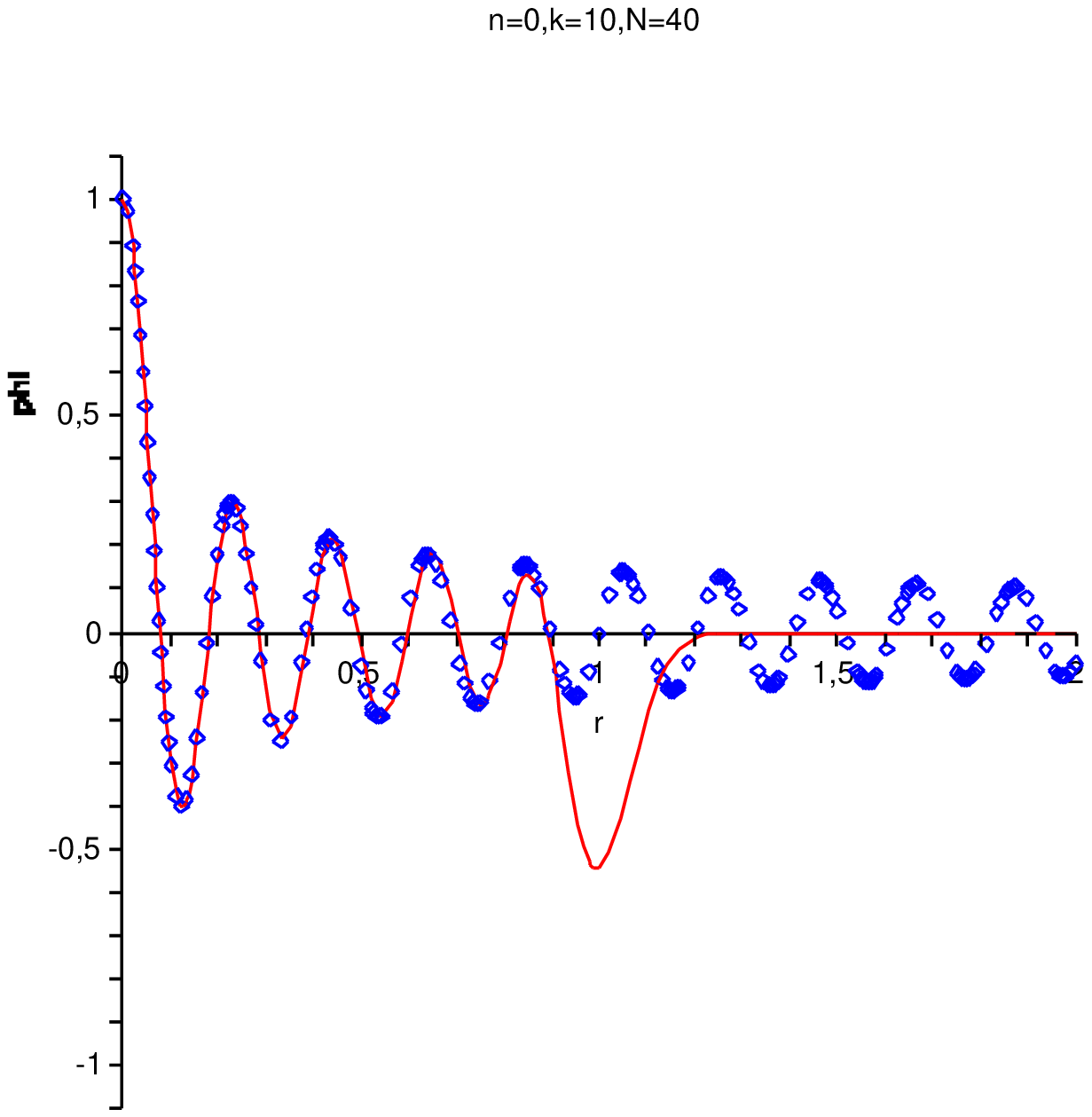}} \caption{\baselineskip=12pt {\it
Comparison of the radial shape for the
$\Phi_{0,10}^{\left(N\right)}$ symbol for $N=30,35,40$.}}
\bigskip
\label{zero_dieci_seconda}
\end{figure}

We now compare the behaviour of the symbols of the fuzzy Bessel
with their ordinary counterparts.
From~\eqn{Phinfuzz} the symbol of a fuzzy Bessel is:
\beq
\Phi_{n,k}^{\left(N\right)}=\left(\frac{\lambda_{n,k}^{\left
(N\right)}}{4}\right)^{n/2}\,r^{n}e^{in\varphi}
e^{-Nr^{2}}\,\sum_{a=0}^{N-n}\,r^{2a}N^{a}\left[\sum_{s=0}^{a}\,
\frac{1}{s!\left(s+n\right)!\left(a-s\right)!}
\left(-\frac{\lambda_{n,k}^{\left(N\right)}}{4N}\right)^{s}\right]\
. \eeq The integer $n$ appears as a phase modulating factor for
the variable $\varphi$. This would be the expansion of the
corresponding Bessel function, where it not for the truncation in
the sum, and the fact that the parameter $\lambda_{0,k}$ has
become the eigenvalue of the fuzzy Laplacian, i.e.\ a solution
of~(\ref{eigeqpolydiago}). For the $n=0$ the expression can be
simplified:
\beqa
\Phi_{0,k}^{\left(N\right)}\left(r\right)&=&
\sum_{a=0}^{N}\,\left(\sum_{s=0}^{a}\,\left(
-\frac{\lambda_{0,k}^{\left(N\right)}} {4N}\right)^{s}
\frac{1}{s!}\left(\begin{array}{c} a \\ s
\end{array}\right)\right)\,e^{-Nr^{2}}r^{2a}\frac{N^{a}}{a!}\nonumber
\\
&=&e^{-Nr^{2}}\,\sum_{a=0}^{N}\,\frac{N^{a}}{a!}\,r^{2a}\,
\mathcal{L}_{\left(a\right)}
\left(\frac{\lambda_{0,k}^{\left(N\right)}} {4N}\right) \ . \eeqa
Where $\mathcal{L}_{\left(a\right)}
\left(\frac{\lambda_{0,k}^{\left(N\right)}} {4N}\right)$ is the
$a^{th}$ Laguerre polynomial in the variable
$\left(\frac{\lambda_{0,k}^{\left(N\right)}} {4N}\right)$. We can
plot the diagonal fuzzy elements. Fig.~\ref{zero_uno} shows that
the zero order fuzzy Bessel state converges to the continuum
eigenfunctions $\Phi_{0,1}\left(r,\varphi\right)$ for values of
$r$ inside the disc of radius $1$, while it converges to zero
outside the disc.
This behaviour is seen to be valid also for eigenstates of
different eigenvalues. The plots for the symbols
$\Phi_{0,2}^{\left(N\right)}$ are in figure~\ref{zero_due}, those
for $\Phi_{0,3}^{\left(N\right)}$ in figure~\ref{zero_tre}.
It is interesting to analyse the fuzzification of
$\Phi_{0,10}\left(r\right)$. The fuzzy symbol is
$\hat{\Phi}_{0,10}^{\left(N\right)}$. For $N=10,20,25$ it is
plotted in figure~(\ref{zero_dieci_prima}).
It is evident that the fuzzy eigenfunction reproduces the
continuum eigenfunction for values of $r$ close to the centre of
the disc, but not on the edge, where a huge bump appears.
In~\cite{beatles} it has been explained that the presence of the
bump on the edge of the disc, in the fuzzification of a function
defined on the plane, is related to the fact that this function
has oscillations of too small wavelenght compared to $\theta$.
This is manifestation of the infrared-ultraviolet mixing
characteristic of noncommutative theories. In the case of
$\Phi_{0,10}\left(r\right)=J_{0}\left(\sqrt{\lambda_{0,10}}r\right)$,
one can immediately see that the oscillation wavelenght of the
continuum eigenfunction is given by $\rho_{\lambda}\,\sim\,1/k\,\sim 1/10$.
In these plots, it is assumed $\theta=1/N$ (the fuzzy disc
truncation), so $\theta$ and $\rho_{\lambda}$ are of compatible
magnitude. In the fuzzy disc limit, $N\,\rightarrow\,\infty$, so
$\theta$ is infinitesimal. The bump disappears, as it is shown in
the other plots (figure~\ref{zero_dieci_seconda}).

\begin{figure}[htbp]
\epsfxsize=2.3 in \centerline{\epsfxsize=2.2
in\epsffile{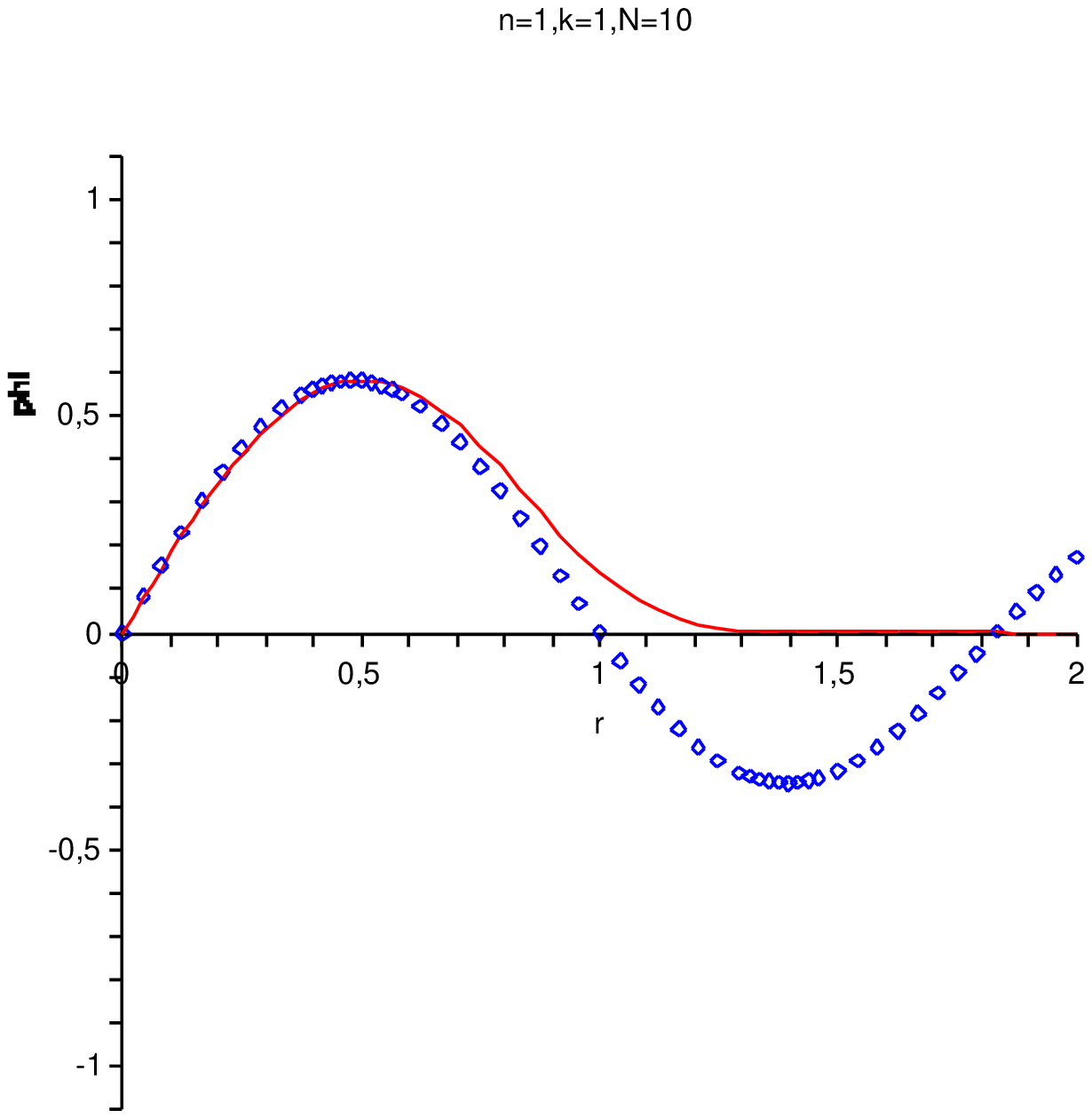}\epsfxsize=2.2
in\epsffile{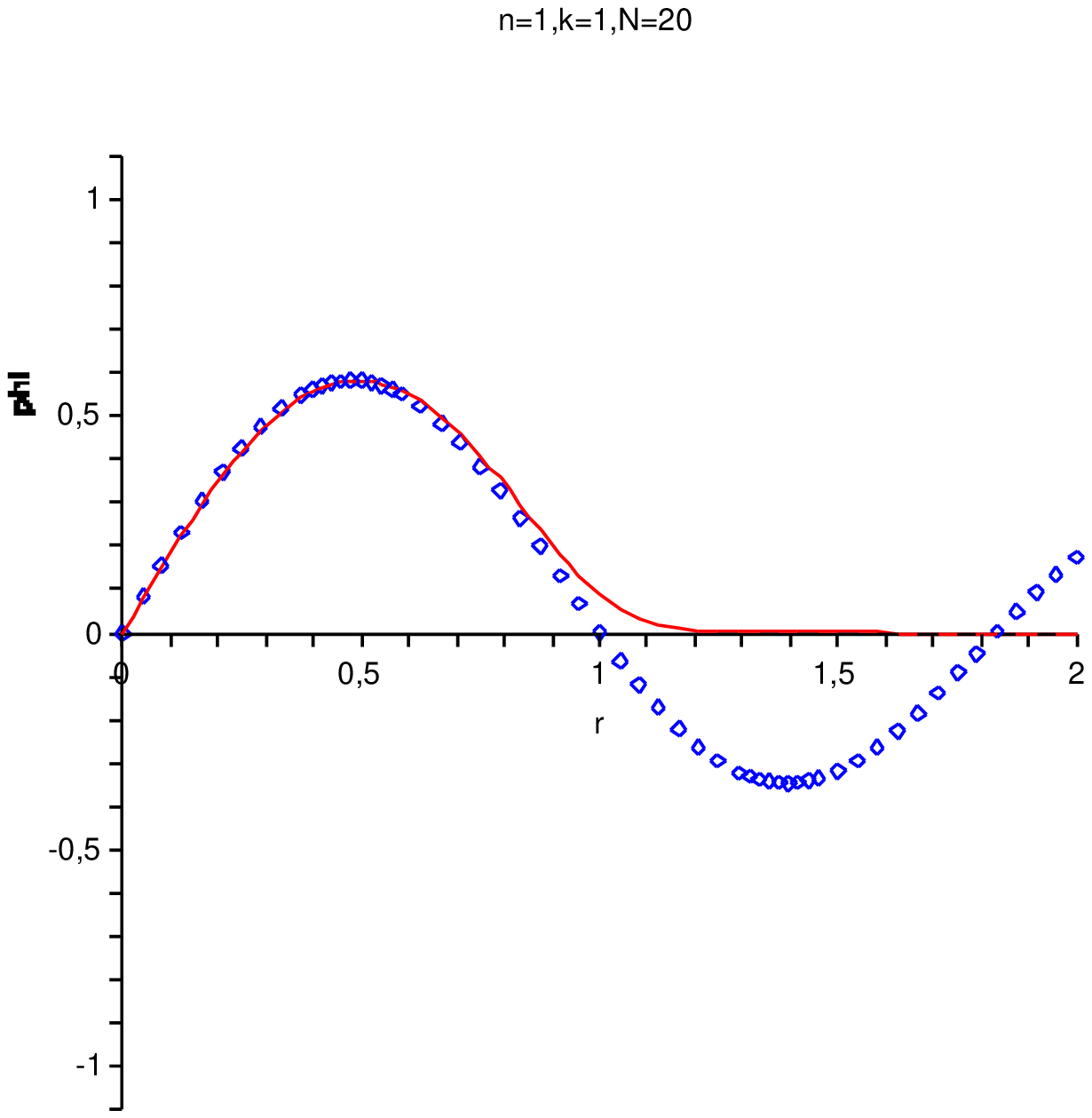}\epsfxsize=2.2
in\epsffile{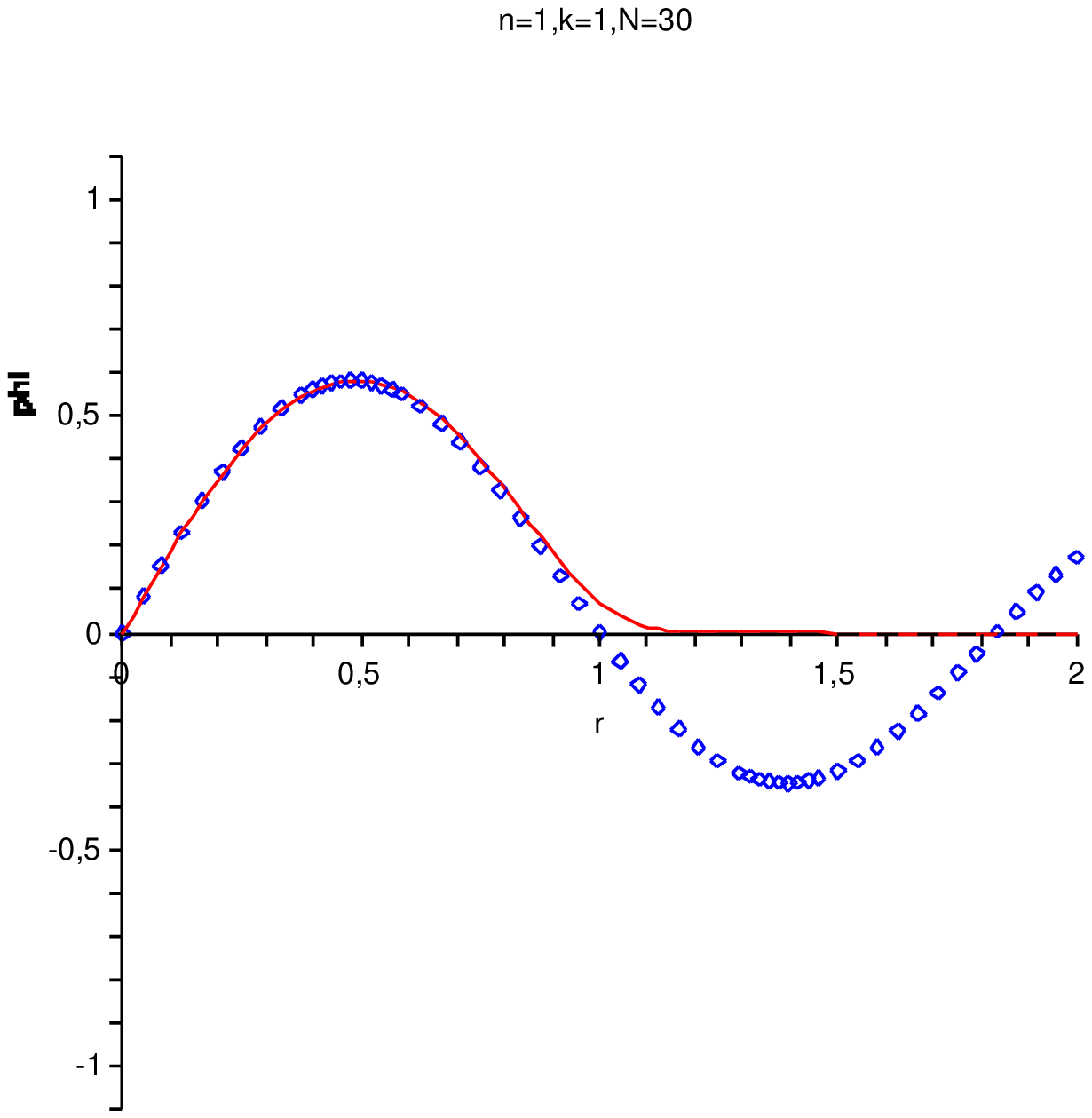}} \caption{\baselineskip=12pt {\it
Comparison of the radial shape of the
$\Phi_{1,1}^{\left(N\right)}\left(r,\varphi\right)$ symbol
(continuum line) for  $N=10,20,30$. }}
\bigskip
\label{uno_uno}
\end{figure}

\begin{figure}[htbp]
\epsfxsize=2.3 in \centerline{\epsfxsize=2.2
in\epsffile{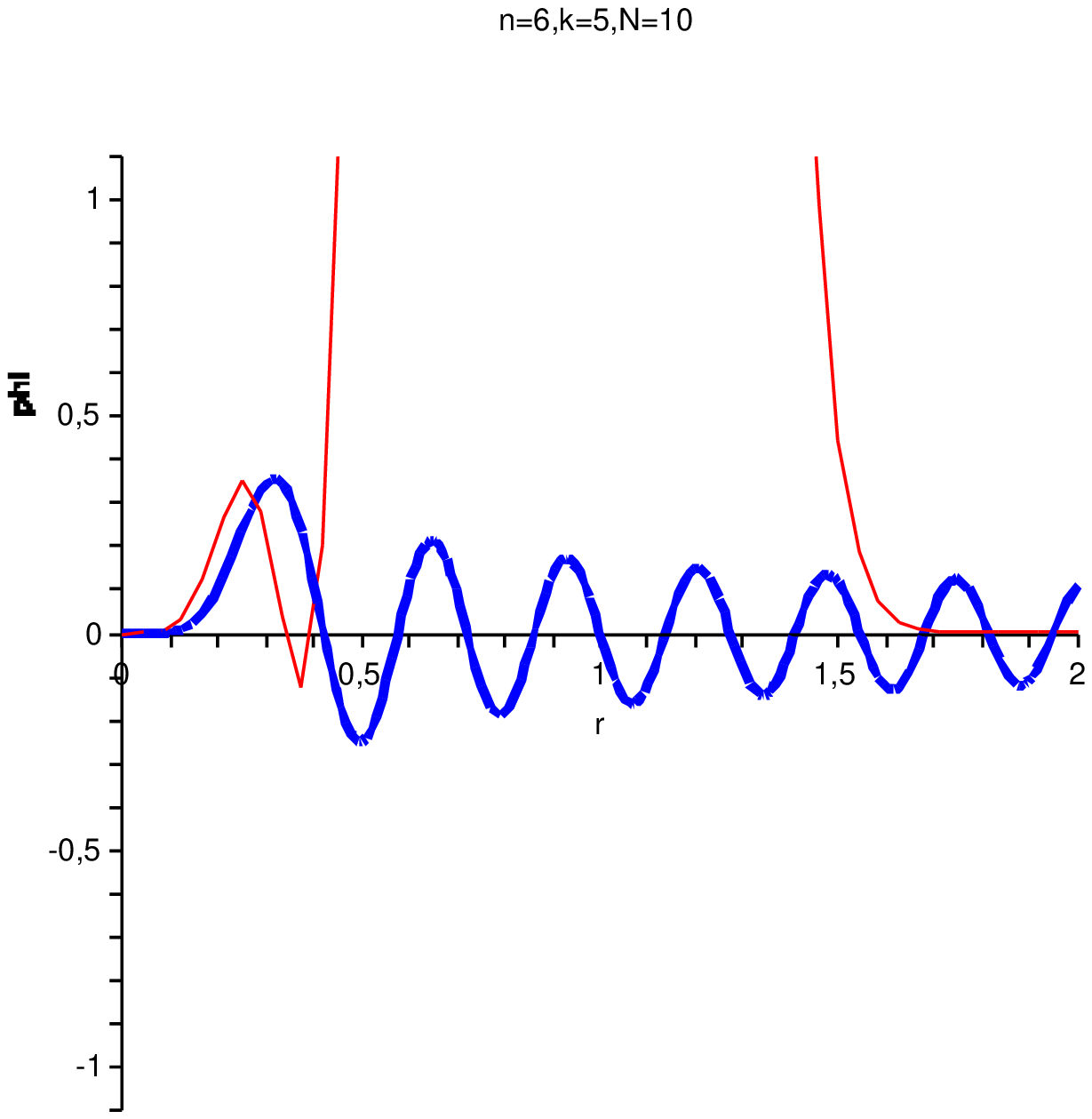}\epsfxsize=2.2
in\epsffile{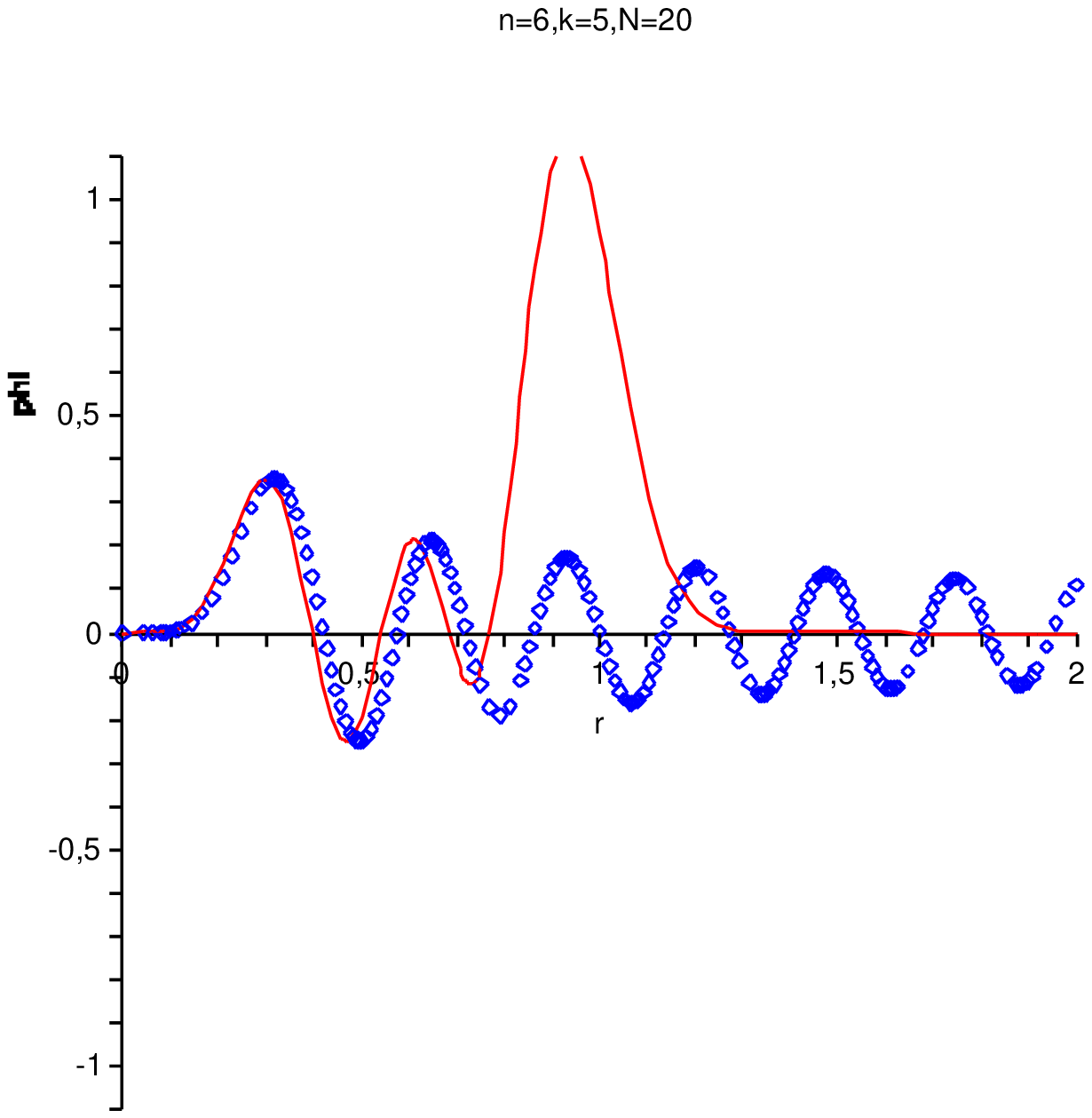}\epsfxsize=2.2
in\epsffile{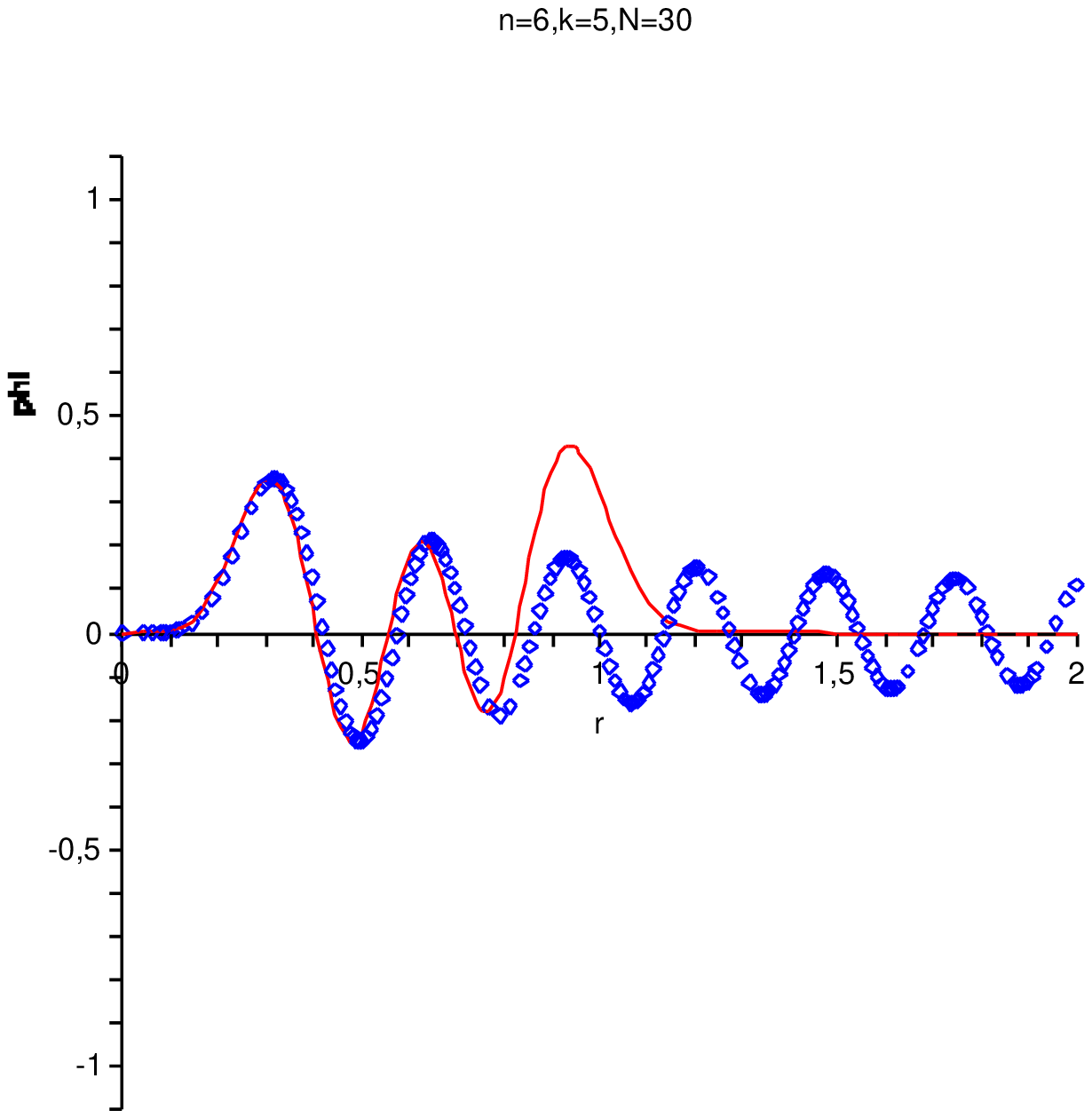}} \caption{\baselineskip=12pt {\it
Comparison of the radial shape of the
$\Phi_{6,5}^{\left(N\right)}\left(r,\varphi\right)$ symbol
(continuum line) for $N=10,20,30$.}}
\bigskip
\label{sei_cinque}
\end{figure}

The non radial functions follow a similar pattern. Their phases
are exactly as the ones of their continuum counterparts, while the
radial parts are similar. A first plot is in Fig.~\ref{uno_uno}, a
second one in Fig.~\ref{sei_cinque}, where the fuzzification
procedure gives again a bump, for small values of $N$. This bump
is seen to disappear in the fuzzy disc limit.

The fuzzy Bessels, being eigenfunctions, provide an efficient way
to calculate a fuzzy Green function, which we calculated
numerically in~\cite{beatles}. Since the fuzzy Laplacian is
represented by an hermitian finite dimensional matrix, its inverse
can be written in terms of a spectral decomposition. If a matrix,
say $M$, is hermitian, then its components satisfy the condition
$M_{ab}=M_{ba}^{*}$ in terms of transposition and complex
conjugation. The eigenvalue problem gives a number of real
eigenvalues equal to the dimension of the space on which the
matrix acts: \beq
\sum_{b}M_{ab}v_{b}^{\left(k\right)}=\lambda^{\left(k\right)
}v_{a}^{\left(k\right)}\ ,
\eeq here $v_{a}^{\left(k\right)}$ indicates the $a^{th}$
components of the eigenvector relative to the eigenvalue
$\lambda^{\left(k\right)}$. The inverse, if it exists, of the
matrix $M$ is a matrix $G$ whose components can be written as:
\beq
G_{sq}=\sum_{k}\,v_{s}^{\left(k\right)}v_{q}^{\left(k\right)}
/\lambda^{\left(k\right)} \ . \eeq If we specialise to the problem
under analysis,  the notion of eigenvector of components
$v_{a}^{\left(N\right)}$ with eigenvalue
$\lambda^{\left(k\right)}$ goes into that of fuzzy Bessel
$\hat{\Phi}^{\left(N\right)}_{n,k}$ matrix, from which it is
immediate to obtain the symbols. The fuzzy Green function becomes:
\beq G^{\left(N\right)}_{\theta}\left(z,z^{\prime}\right)=
\sum_{n=-N}^{+N}\sum_{k=1}^{N+1-\left|n\right|}
\frac{\Phi_{n,k}^{\left(N\right)}\left(z^{\prime}\right)^{*}
\Phi_{n,k}^{\left(N\right)}\left(z\right)}
{\lambda^{\left(N\right)}_{|n|,k}}\ . \eeq

Since the fuzzy Bessels play a role similar to fuzzy harmonics for
the fuzzy sphere algebra, we can now make describe the process of
approximating the algebra of functions on a disc with matrices
more precise. In complete analogy with~\eqn{harmexp}
and~\eqn{fuzzharmexp}, if $f$ is square integrable with respect to
the standard measure on the disc $d\Omega= r dr d\phi$, it can be
expanded in terms of Bessel functions:
\beq
f\left(r,\varphi\right)=\sum_{n=-\infty}^{+\infty}\,\sum_{k=1}^{\infty}\,
f_{nk}e^{in\varphi}J_{\left|n\right|}
\left(\sqrt{\lambda_{\left|n\right|,k}}r\right) \eeq and it is
possible to truncate: \beq
f^{\left(N\right)}\left(r,\varphi\right)=\sum_{n=-N}^{+N}\sum_{k=1}^{N+1-|
n|}\,f_{nk}e^{in\varphi}J_{\left|n\right|}\left(
\sqrt{\lambda_{\left|n\right|,k}}r\right)
=\sum_{n=-N}^{+N}\sum_{k=1}^{N+1-|
n|}\,f_{nk}\,\Phi_{n,k}\left(r,\varphi\right) \ . \eeq This  set
of functions  is a vector space, but it is no more an algebra,
with the standard definition of sum and pointwise product,  as the
product of two truncated functions will get out of the algebra.
The mapping from truncated functions into finite rank matrices:

\beq 
f^{(N)}\rightarrow \hat{f}^{\left(N\right)}_{\theta}=
\sum_{n=-N}^{+N}\sum_{k=1}^{N+1-| n|}\,f_{nk}
\hat{\Phi}_{n,k}^{\left(N\right)} 
\eeq 
endows the set of functions
with a noncommutative product, inherited from the matrix product,
which makes it into a non abelian algebra. The  formal limit
$N\rightarrow \infty$ with the constraint $N\theta=1$ is the
abelian algebra of functions on the disc. The sequence of
nonabelian algebras $\mathcal{A}_\theta^{(N)}$ is what we call the
\emph{fuzzy disc}.

\setcounter{section}{0}
\renewcommand{\thesection}{\Alph{section}}
\setcounter{equation}{0}

\section{Weyl-Wigner Formalism \label{appWW}}
\setcounter{equation}{0}

In this appendix we give a very short introduction to the
Weyl-Wigner formalism used to define maps between functions on a
``classical" plane and a set of operators on a Hilbert
space~\cite{wignerreview}.

Given a symplectic real, finite dimensional, vector space
$\left(L,\omega\right)$, where the symplectic form $\omega$ is
translationally invariant, i.e.\ with constant coefficients, the
set of Displacement operators define a Weyl system, a unitary
representation of canonical commutation relations in the
exponentiated unitary version~\cite{baezsz}:
\beqa
\Do\,&:&\,L\,\mapsto\,\mathcal{U}\left(\mathcal{H}\right)
\nonumber\\
\Do\left(z+u\right)&=&
e^{i\omega\left(z,u\right)/2\theta}\Do\left(z\right)\Do\left(u\right)
\ . \eeqa Here $\theta$ is a parameter: the Displacement operators
can be seen to set a unitary projective representation of the
translation group, where the phase factor is related to the
sympletic structure defined on the vector space $L$. These
operators also define a representation of the Heisenberg-Weyl
group~\cite{folland}. On every one dimensional subspace of $L$
($\alpha$ is a real scalar), via the Stone-von Neumann theorem:
\beq
\Do\left(\alpha z\right)=e^{i\alpha\Go\left(z\right)\theta}
\eeq
and these generators satisfy:
\beq
\left[\Go\left(z\right),\Go\left(u\right)\right]\,=
\,i\theta\omega\left(z,u\right)\,\id \ . \eeq This shows that a
Weyl system is a way to formalise a set of canonical commutation
relations for quantum observables, recovered as hermitian
generators of a unitary ray representation. In this perspective
$\theta$ acquires the role of a noncommutativity parameter. In the
standard approach, the symplectic vector space is seen as a phase
space for the classical dynamics of point particle: generators of
a Weyl system represent the position and momentum observable for a
quantum dynamics of point particle.

Using the definition of a Weyl system, it is possible to define a
Weyl map, that is a map from functions on the vector space
$\left(L\omega\right)$ (dim $L=2n$) to $Op\left(\hil\right)$, the
set of operators on the Hilbert space on which Displacements are
realised:
\beq
\hat{f}=\int\,\frac{dw}{\left(2\pi\theta\right)^{n}}
\,\tilde{f}\left(w\right)\Do\left(w\right)\ ,
\eeq
where:
\beq
\tilde{f}\left(w\right)=\int
\frac{dz}{\left(2\pi\theta\right)^{n}}\,f\left(z\right)\,
e^{-i\omega\left(z,w\right)/\theta}
\eeq
is the symplectic Fourier transform~\cite{folland} of the function
$f$. The Weyl map can be seen as a quantization map for classical
observables defined on a phase space: it can be inverted by the
Wigner map:
\beq
\tilde{f}\left(w\right)=Tr\left[\hat{f}\Do^{\dagger}\left(w\right)\right]
\ . \eeq So the Weyl-Wigner maps define a bijection between a set
of classical observables and a set of quantum observables. This is
the building block of a formalism suited to study both the problem
of quantizing a classical system and performing a classical limit
of a quantum system.

The standard Weyl-Wigner maps can be modified by the introduction
of a term, called weight:
\beqa
\hat{f}_{\lambda}\,&=&\,\int\,\frac{dw}{\left(2\pi\theta\right)^{n}}
\,\tilde{f}\left(w\right)\lambda\left(w\right)\Do\left(w\right)
\nonumber
\\
\tilde{f}\left(w\right)&=&
\lambda\left(w\right)^{-1}Tr\left[\hat{f}_{\lambda}
\Do^{\dagger}\left(w\right)\right] \ . \eeqa This weight can be
proved to be related to a more general class of ray representation
of the translation group, and is seen to be responsible both for a
specific "ordering" in the quantization procedure and a definition
of a specific domain of mathematical applicability of the
bijection, as stressed in the main text.

\section{Generalised coherent states}\label{appz}
\setcounter{equation}{0}

In this paper the concept of generalised coherent states has been
extensively used.  The aim of this appendix is to briefly
recollect the main definitions and results, to fix notations and
facilitate the reading. The main reference for coherent states
is~\cite{perelomov}.

Consider a Lie group $G$, and $\Uo\left(g\right)$ a unitary
irreducible representation of this group on a Hilbert space
$\hil$. Chosen a \emph{fiducial} vector $|\psi_{0}\ra$ in $\hil$,
one obtains a set of vectors for each element of the group, acting
on it with $\Uo\left(g\right)$:
\beq
|\psi_{g}\ra\,\equiv\,\Uo\left(g\right)|\psi_{0}\ra \ . \eeq Two
such vectors are considered \emph{equivalent} if they correspond,
quantum-mechanically, to the same state, i.e. if they differ by a
phase. So $|\psi_{g}\ra\,\simeq\,|\psi_{g^{\prime}}\ra$ if
$|\psi_{g}\ra\,=e^{i\phi\left(g,g^{\prime}\right)}
\,|\psi_{g^{\prime}}\ra$. This condition is equivalent to
$\Uo\left(g^{\prime -1}g\right)|\psi_{0}\ra\,=
e^{i\phi\left(g,g^{\prime}\right)}\,|\psi_{0}\ra$. If $H$ is the
subgroup of $G$ whose elements are represented, by $\Uo$, as
operators whose action on the fiducial vector is just a
multiplication by a phase, then the equivalence relation is among
points of $G$, and the quotient is the space $G/H$. If $H$ is
maximal, then it is called isotropy subgroup for the state
$|\psi_{0}\ra$. Choosing a representative $g\left(x\right)$ in
each equivalence class $x\,\in\,X=G/H$ (which is a cross section
of the fiber bundle $G$ with base $X$) defines a set of vectors on
$\hil$, depending, clearly, on $G$ and $|\psi_{0}\ra$. This set of
states is called a \emph{system of coherent states} for $G$. The
state corresponding to the vector $| x\ra$ may be considered as
the range of a rank one projector in $\hil$. Thus, the system of
generalised coherent states determines a set of one dimensional
subspaces in $\hil$, parametrised by points of the homogeneous
space $X=G/H$. An evolution of this analysis drives naturally to
the issue of overcompleteness for the system of coherent states,
mentioned in~(\ref{HWcomplete}),~(\ref{SU2complete}).

\section{Explicit Calculations for the Eigenvalues \label{appdiagospectral}}
The aim of this appendix is to show an explicit proof of the
results claimed in the main text about the eigenvalue problem for
the fuzzy Laplacian. In particular, we want to prove that the
equation~(\ref{eigeqENn}) is equivalent to the
equation~(\ref{eigeqpolyn}). From this equivalence, it will be
proved also the equivalence between equations~(\ref{eigeqdiagoEN})
and equation~(\ref{eigeqpolydiago}), as they are a special case of
the first two for $n=0$.

The polynomial $E_{N}^{\left(n\right)}$ is given, identifying
$\lambda^{\left(N\right)}_{n,k}/4N=x^{N}$, by~(\ref{eqENn}): \beq
E_{N}^{\left(n\right)}\left(x_N\right)=
\sum_{k=0}^{N-n+1}\,\left(-1\right)^{k+N-n+1}\,
\frac{\left(N+1\right)!\left(N-n+1\right)!}
{k!\left(n+k\right)!\left(N-n-k+1\right)!}\,\left(x_N\right)^{k}=0
\label{appeqENn}\ . \eeq The characteristic polynomial
$P^{\left(n\right)}_{N}$ is given by the determinant of the
matrix~(\ref{secnmat}) given recursively by~\eqn{detn}. To prove
the equality of the two polynomials, the first step is to prove
that $E_{N}^{\left(n\right)}$ satisfies the same recursive
relation that $P_{N}^{\left(n\right)}$ does.

Considering the quantity: \beqa
&{}&\left(x_N-\left(2N-n+1\right)\right)\,
E_{N-1}^{\left(n\right)}\left(x_N\right)
\,-\,N\left(N-n\right)\,E_{N-2}^{\left(n\right)}\left(x_N\right)
=\nonumber \\
&{}&=\sum_{k=0}^{N-n}\left(-1\right)^{k+N-n}\frac{N!\left(N-n\right)!}
{k!\left(n+k\right)!\left(N-n-k\right)!}\,\left(x_N\right)^{k+1}\,+\nonumber
\\
&{}& -\sum_{k=0}^{N-n}\left(-1\right)^{k+N-n}\left(2N-n+1\right)
\frac{N!\left(N-n\right)!}
{k!\left(n+k\right)!\left(N-n-k\right)!}\,\left(x_N\right)^{k}\,+\nonumber
\\ &{}&-
\sum_{k=0}^{N-n-1}\left(-1\right)^{k+N-n+1}\frac{N!\left(N-n\right)!}
{k!\left(n+k\right)!\left(N-n-k-1\right)!}\,
\left(x_N\right)^{k}\label{ENnutinam}\ . \eeqa Where the
coefficient of the zeroth order term is:
\beq
\left(\left(-1\right)^{N-n+1}\frac{N!}{n!}
\left(2N-n+1\right)-\left(-1\right)^{N-n+1}\frac{N!}{n!}\left(N-n\right)\right)
=\left(-1\right)^{N-n+1}\frac{\left(N+1\right)!}{n!}\ . \eeq It
coincides with the coefficient of the zeroth order term of the
polynomial $E^{\left(n\right)}_{N}$.

Let us consider, in the expression~(\ref{ENnutinam}), the
coefficient of the $q^{th}$ order term, with $1\leq q\leq N-n-1$.
It is equal to:
\beqa
&\left(-1\right)^{N-n+q-1}N!\left(N-n\right)!\left(
\frac{1}{\left(a-1\right)!\left(n+a-1\right)!\left(N-n-a+1\right)!}+
\right.&\\&\left.
\frac{2N-n+1}{a!\left(n+a\right)!\left(N-n-a\right)!}-
\frac{1}{a!\left(n+a\right)!\left(N-n-a-1\right)!}\right)\left(x_N\right)^{q}=
\left(-1\right)^{N-n+a+1}\frac{\left(N+1\right)!\left(N-n+1\right)!}
{a!\left(a+n\right)!\left(N-n-a+1\right)!}
\left(x_N\right)^{q}&\nonumber \ . \eeqa This coefficient is equal
to the coefficient of the $q^{th}$ order term in
$E_{N}^{\left(n\right)}$. The equality of the coefficients, of the
remaining order terms, of the polynomial (\ref{ENnutinam}) with
those of the polynomial $E^{\left(n\right)}_{N}\left(x_N\right)$
can be easily checked.

At this stage, we have proved that the polynomial
$E^{\left(n\right)}_{N}\left(x_N\right)$ satisfies the same
recursive relation that the polynomial
$P^{\left(n\right)}_{N}\left(x_N\right)$ does satisfy. The
recursive relation they satisfy is such that the $N^{th}$ term of
the sequence $E_{N}^{\left(n\right)}$, for each fixed $n$, depends
on both the $\left(N-1\right)^{th}$ and $\left(N-2\right)^{th}$
terms of the sequence. To prove the complete equivalence of these
two polynomials, we thus need to prove that they explicitly
coincide at a pair of consecutive steps.

In the main text it has been stressed that there is a constraint
for allowed positive $n$, namely an upper bound: $n\,\leq\,N+1$.
This means that, for a fixed $n$, that is for a fixed subspace in
the fuzzy algebra, $N\,\geq\,n+1$. For $N=n-1$ both polynomials
$E_{n-1}^{\left(n\right)}\left(x_N\right)$ and
$P_{n-1}^{\left(n\right)}\left(x_N\right)$ are trivial, so we
study at first the case of $N=n$. We have:
\beq
E_{n}^{\left(n\right)}=\sum_{k=0}^{1}\,\left(-1\right)^{k+1}\,
\frac{\left(n+1\right)!}{k!\left(n+k\right)!\left(1-k\right)!}\,
\left(x^{\left(N\right)}\right)^{k}=x^{\left(N\right)}-\left(n+1\right)\eeq
while the matrix representing the action of the fuzzy Laplacian is
one-dimensional, so that: \beq
P_{n}^{\left(N\right)}=x_N-\left(2n+1\right)\ . \eeq The second
case is for $N=n+1$. We have: \beqa
E_{n}^{\left(n+1\right)}&=&\sum_{k=0}^{2}\,\left(-1\right)^{k+2}\,
\frac{2\left(n+2\right)!}{k!\left(n+k\right)!\left(2-k\right)!}\,
\left(x_N\right)^{k}\nonumber \\
&=&\,\left(x_N\right)^{2}-2\left(n+2\right)x_N+
\left(n+1\right)\left(n+2\right)\ . \eeqa The matrix representing
the action of the fuzzy Laplacian is now two-dimensional, so that
the characteristic polynomial is: \beqa
P_{n+1}^{\left(n\right)}&=& \det \left(\begin{array}{cc}
x_N-\left(n+1\right) & \sqrt{n+1}
\\ \sqrt{n+1} & x_N-\left(n+3\right) \end{array}
\right) \,=\nonumber \\ &=& \left(x_N\right)^{2}-2\left(n+2\right)
x_N+\left(n+1\right)\left(n+2\right)\ . \eeqa This proves the
equivalence of the two polynomials.

\subsection*{Acknowledgments}
We thank  D.~Bercioux, P.~Lucignano, P.~Santorelli and A.~Stern
for help at various stages. This work has been supported in part
by the {\sl Progetto di Ricerca di Interesse Nazionale {\em
SInteSi}}.

\end{document}